\documentclass[sigconf]{acmart}

\usepackage{url}
\usepackage{enumerate}
\usepackage{subfig}
\usepackage{multirow}
\usepackage{tabularx}
\usepackage{color}

\usepackage[capitalise]{cleveref}
\AtBeginDocument{%
  }

\setcopyright{acmcopyright}
\copyrightyear{2023}
\acmYear{2023}
\acmDOI{XXXXXXX.XXXXXXX}

\acmConference[The Web Conference]{Conference}{XXXXXX}{XXXXXX}
\acmPrice{15.00}
\acmISBN{978-1-4503-XXXX-X/18/06}

\begin{document}

\title{PerCoNet: News Recommendation with Explicit Persona and Contrastive Learning}

\author{Rui Liu}
\email{lr@buaa.edu.cn}
\authornotemark[1]
\affiliation{%
  \institution{Beihang University}
  \city{Beijing}
  \state{100191}
  \country{China}
}

\author{Bin Yin}
\email{ZY2006163@buaa.edu.cn}
\authornotemark[2]
\affiliation{%
	\institution{Beihang University}
	\city{Beijing}
	\state{100191}
	\country{China}
}

\author{Ziyi Cao}
\email{muhhpu@gmail.com}
\authornotemark[3]
\affiliation{%
	\institution{Beihang University}
	\city{Beijing}
	\state{100191}
	\country{China}
}

\author{Qianchen Xia}
\email{qianchenxia@buaa.edu.cn}
\authornotemark[4]
\affiliation{%
	\institution{Tsinghua University}
	\city{Beijing}
	\state{100091}
	\country{China}
}

\author{Yong Chen}
\authornote{If you have any doubts, please feel free to contact me.}
\email{alphawolf.chen@gmail.com}
\authornotemark[5]
\affiliation{%
	\institution{Beijing University of Posts and Telecommunications}
	\city{Beijing}
	\state{100876}
	\country{China}
}

\author{Dell Zhang}
\email{dell.z@ieee.org}
\authornotemark[6]
\affiliation{%
	\institution{Thomson Reuters Labs}
	\city{London}
	\state{}
	\country{the United Kingdom}
}




\begin{abstract}
Personalized news recommender systems help users quickly find content of their interests from the sea of information.
Today, the mainstream technology for personalized news recommendation is based on deep neural networks that can accurately model the semantic match between news items and users' interests. 
In this paper, we present \textbf{PerCoNet}, a novel deep learning approach to personalized news recommendation which features two new findings:
(i) representing users through \emph{explicit persona analysis} based on the prominent entities in their recent news reading history could be more effective than latent persona analysis employed by most existing work, with a side benefit of enhanced explainability;
(ii) utilizing the title and abstract of each news item via cross-view \emph{contrastive learning} would work better than just combining them directly. 
Extensive experiments on two real-world news datasets clearly show the superior performance of our proposed approach in comparison with current state-of-the-art techniques.
\end{abstract}



\maketitle

\section{Introduction}

Thanks to online news services (such as those provided by Google\footnote{\url{https://news.google.com/}}, MSN\footnote{\url{https://www.msn.com/en-us/news}}, and Toutiao\footnote{\url{https://www.toutiao.com/}}), nowadays people can easily stay on top of what is happening in the whole wide world without leaving home.
With so many news items collected every day, these news platforms need to provide \emph{personalized news recommendations}~\cite{Hao-Fatigue-WWW-2016,Jianxun-MultiChannel-IJCAI-2018,Yong-HyperNews-IJCAI-2020,Jingwei-Federated-EMNLP-2021,Shansan-Session-SIGIR-2022,Seonghwan-Diversity-WWW-2022} to make it easier for users to find content that matches their preferences.

\begin{figure}[!tb]
	\centering
	\includegraphics[width=0.48\textwidth]{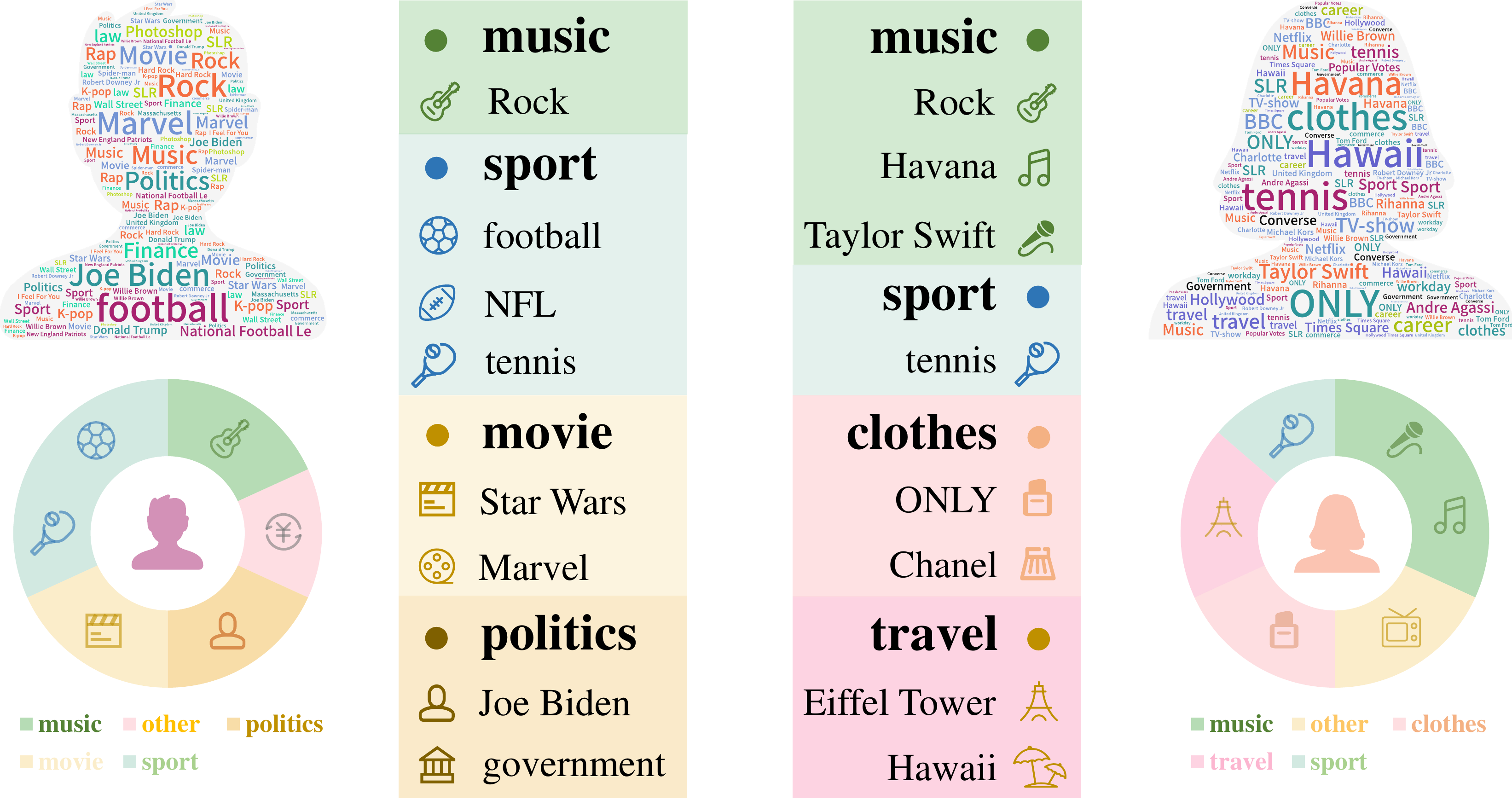}
	\caption{An illustration of different users' explicit personae.}
	\label{fig:Example}
\end{figure}

Existing personalized news recommendation methods usually recommend news items to a target user based on the semantic match between the textual content of candidate news items and the user interest inferred from their previous news reading behaviors. 
For example, 
DKN~\cite{Hongwei-DKN-WWW-2018} designs a multi-channel and word-entity-aligned knowledge-aware CNN to represent candidate news items and also builds an attention module to dynamically aggregate a user's recent browsed news items. 
DAN~\cite{Qiannan-DAN-AAAI-2019} utilizes PCNN to encode each candidate news item with its title and profile, and develops ARNN/ANN to extract a user's current interests. 
LSTUR~\cite{Mingxiao-LSTUR-ACL-2019} leverages an attention network to learn informative representation from news titles and categories, and a GRU network to learn long- and short-term representations from user's recently browsed news items.
FIM~\cite{Heyuan-FIM-ACL-2020} captures more fine-grained interest matching signals by analyzing the interactions between each pair of news items at multiple levels of semantic granularity.
UNBERT~\cite{Qi-UNBERT-IJCAI-2021} induces a pre-trained model to alleviate the cold-start problem, and performs multi-grained user-news matching at both word-level and news-level via WLM and NLM.
GREP~\cite{Zhaopeng-GREP-TKDD-2022} borrows the multi-head attention mechanism from NRMS~\cite{Chuhan-NRMS-EMNLP-2019} to encode the sentence-level semantic representation of the news title, and utilizes three modules (namely multi-head self-attention mechanism, graph transformer, and bi-directional interaction) to encode users' existing and potential interests.
More examples in this line of research include NRMS~\cite{Chuhan-NRMS-EMNLP-2019}, KRED~\cite{Danyang-KRED-RecSys-2020}, User-as-Graph~\cite{Chuhan-User-as-Graph-IJCAI-2021}, and
MM/Feed-Rec~\cite{Chuhan-MMRec-SIGIR-2022,Chuhan-FeedRec-WWW-2022}.

Can we model users' interests explicitly with their relevant entities (in addition to implicitly inferring from their past behaviors), and thus obtain better news/user encoders for personalized recommendation? 
Motivated by this question, we construct an \emph{explicit persona} for each user and then embed them into the classic news and user encoders with an interactive attention mechanism. 
Here a user's explicit persona refers to a set of \emph{entities} that can reflect their news reading interests, as illustrated in \cref{fig:Example}. 

Moreover, some recent studies in this area like 
CPRS~\cite{Chuhan-CPRS-IJCAI-2020},
HyperNews~\cite{Yong-HyperNews-IJCAI-2020}, 
KRED~\cite{Danyang-KRED-RecSys-2020},
PENR~\cite{Jingkun-PENR-CIKM-2021},
and MINER~\cite{Jian-Miner-ACL-2022}
have discovered that the task of personalized news recommendation can be reinforced by some related tasks such as `category classification', `active-time prediction', `satisfaction predictor', `popularity prediction', and `local news detection' within a \emph{multi-task learning} framework. 
Inspired by this observation, we propose a new related task cross-view contrastive learning which compares different user encoder's outputs over cross-view inputs (`abstracts' v.s. `titles'), so as to further enhance the shared news and user encoders' modeling capabilities.

In summary, what we propose in this paper is a novel deep neural \underline{net}work with explicit \underline{per}sona and \underline{co}ntrastive learning, dubbed \textbf{PerCoNet}.
It is able to make better personalized news recommendations by learning better representations for the target user's preferences as well as the candidate news item's semantics, as demonstrated by our extensive experiments on two real-world news datasets.
The main contributions of this work are two new findings that have never been reported before in the research literature on personalized news recommendation:
\begin{enumerate}[(i)]
	\item representing users through \emph{explicit persona analysis} based on the prominent entities in their recent news reading history could be more effective than latent persona analysis employed by most existing work, with a side benefit of enhanced explainability;
	\item utilizing the title and abstract of each news item via cross-view \emph{contrastive learning} would work better than just combining them directly.
\end{enumerate}

\section{Related Work}

Our PerCoNet is mostly related to the research fields of news recommendation and contrastive learning, as detailed below.

\subsection{News Recommendation}

Personalized news recommendation can effectively improve users' reading experience and therefore has been widely studied in both academia and industry. 
Existing methods usually model news items (based on their text content) and user interests (based on their reading history) independently, and then match news items with users according to the similarity between their embeddings. 
For example, 
NPA~\cite{Chuhan-NPA-KDD-2019} learns news representations from the news titles using a CNN, and learns user representations from each user's clicked news items via personalized word- and news-level attention mechanisms.
CUPMAR~\cite{Dai-CUPMAR-WISE-2021} leverages multiple properties of each news item with advanced neural network layers to derive news representations, and extracts users' long-term and recent preferences from their news reading history to derive user representations.
User-as-Graph~\cite{Chuhan-User-as-Graph-IJCAI-2021} represents each news item through multi-view attention over its title, topic category, and entities, and represents each user as a personalized heterogeneous graph to better model the relatedness between user behaviors.	
To sum up, existing approaches to personalized news recommendation like those mentioned above all try to employ various techniques (such as GRU~\cite{Mingxiao-LSTUR-ACL-2019,Rongyao-MINS-ICASSP-2022}, GNN~\cite{Yu-ReGCN-SIGIR-2021,Linmei-GNN-ACL-2020,Zhaopeng-GREP-TKDD-2022}, knowledge graph~\cite{Hongwei-DKN-WWW-2018,Shaoyun-WG4Rec-CIKM-2021,Danyang-anchorGK-KDD-2021,Kevin-KG-WWW-2019}, and attention mechanism~\cite{Chuhan-NRMS-EMNLP-2019,Lingkang-DCAN-DASFAA-2021,Yong-HyperNews-IJCAI-2020}) to infer latent user interests from the news items that the user has read. 
Different from such an implicit learning manner, our proposed PerCoNet method builds an explicit persona for each user and injects it into the deep user/news encoders, which leads to higher recommendation accuracy.
This resembles the well-known technique \emph{explicit semantic analysis (ESA)}~\cite{gabrilovich2006overcoming,egozi2011concept} which, in contrast to \emph{latent semantic analysis (LSA)}, defines concepts using the documents in an external corpus (like the English Wikipedia) as human-readable semantic units. 
Similarly, PerCoNet tries to describe users' personae using the external set of entities appeared in their news reading history as explainable semantic units.


\subsection{Contrastive Learning}

One of the most powerful techniques in self-supervised learning is \emph{contrastive learning}~\cite{Yonglong-goodCL-NeurIPS-2020,Ching-Revisiting-ICML-2022,Tiening-Rumor-WWW-2022,Mohammadreza-CrossCLR-ICCV-2021,Thomas-WorldModels-ICLR-2020,Senthil-DemystifyingCL-NeurIPS-2020,Jizong-SelfPaced-NeurIPS-2021} which tries to learn without labels such an embedding space that the views of the same data instance (i.e., positive examples) have more similar representations than the views of different data instances (i.e., negative examples).
Recently contrastive learning has achieved wide-spread successes.
In the field of computer vision, SimCLR~\cite{Ting-SimCLE-ICML-2020} and MoCo~\cite{Kaiming-Moco-CVPR-2020} regard the augmentations to the same image as positive examples and the augmentations to different ones as negative examples.
As for natural language processing tasks, SimCSE~\cite{Tianyu-SimCSE-EMNLP-2021} considers the representations of the same sentence obtained from random dropouts as positive examples, and DeCLUTR~\cite{John-DeCLUTR-ACL/IJCNLP-2021} randomly select textual segments from the same document as positive examples.

Some recommender systems have started to adopt contrastive learning in their machine learning pipelines.
For instance, 
DHCN~\cite{Xin-DHCN-AAAI-2021} 
learns node representations with contrastive learning on different hypergraphs;
RAP~\cite{Xiaohai-RAP-IJCAI-2021} presents a contrastive learning based sequence denoising model, which produces positive and negative sub-sequences by the process of sequence denoising.
However, so far most recommendation methods with contrastive learning focus on graph network model and sequence model, and they generate positive and negative samples by masking or deleting the relations between graph or sequence structures.
With regard to news recommendation, there exist rich textual content attributes such as titles and abstracts which naturally lend themselves to contrastive learning as two different views of each news item.
In our PerCoNet model, a new cross-view contrastive learning module is introduced to enhance news and user encoders' modeling capabilities, and thus elevate the performance of news recommendation.

\section{Proposed Method}

\begin{figure*}[!tb]
	\centering
	\includegraphics[width=0.99\textwidth]{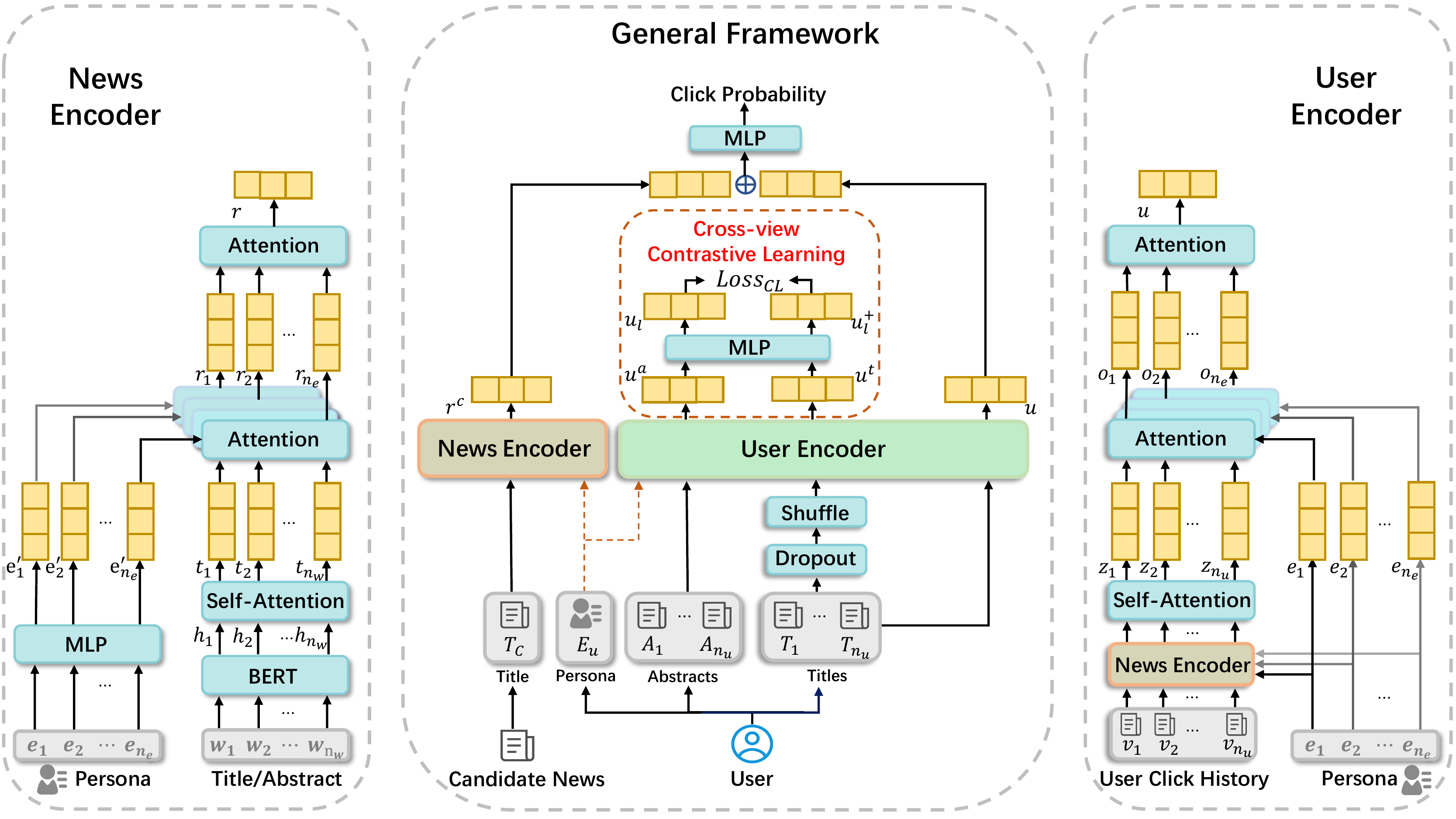}
	\caption{The neural architecture of our proposed PerCoNet model.}
	\label{fig:PerCoNet}
\end{figure*}

In this section, we elaborate on our proposed approach to personalized news recommendation that consists of five stages.
First, we discuss how to build an explicit persona for each user.
Second, we describe the news encoder based on the news item's textual content (i.e., title and abstract) and the associated personae information. 
Third, we present the user encoder based on the user's recently-read news items as well as the user's persona. 
Next, we talk about the task of click-probability prediction utilizing the news and user embeddings.
Finally, we detail a novel cross-view contrastive learning sub-network.
The neural architecture of our proposed PerCoNet is shown in \cref{fig:PerCoNet}.

\subsection{Explicit Persona Construction}

The persona $\mathbf{E}_u$ of a user $u$ is constructed as a set of entities to explicitly characterize $u$'s news reading interests, i.e., $\mathbf{E}_u = \{\mathbf{e}_1,\mathbf{e}_2,\cdots,\mathbf{e}_{n_e}\}$, where $\mathbf{e}_i$ is the $i$-th entity and $n_e$ is the number of entities in the persona.
\cref{fig:Example} illustrates that such explicit personae reveal each user's interests in a human-readable fashion.
For example, Jobs (left) enjoys movies and politics (e.g., `Star Wars' \& `Joe Biden'), while Amy (right) likes clothes and travel (e.g., `Chanel' \& `Hawaii'); both of them are fond of music and sport (e.g., `rock' \& `tennis'). 

In many news datasets, a news item often comes with a pre-configured field that contains the prominent entities in that article.
Therefore we can extract such entities from a user's recently read news items to form his/her explicit persona. 
Specifically, for each user, we locate the most recent $G$ (e.g., $20$) read news items each with the top $K$ (e.g., $4$) entities, and then pool them together to construct the final persona.

When there are external knowledge graphs~\cite{Hongwei-DKN-WWW-2018} or social networks~\cite{Gabriella-profile-SIGIR-2016} available, the users' personae could be further enriched with these extra information.
In this paper, we focus on exploiting the entities (self-contained in public datasets) for a fair comparison with other alternative methods in our experimental evaluation.

\subsection{News Encoder}
The news encoder is devised to learn news representations from news titles/abstracts as well as personae. 
The corresponding sub-network module is shown on the left side of \cref{fig:PerCoNet}.

For a given news item $v$, a pre-trained BERT model~\cite{Jacob-BERT-NAACL-2019} is employed to generate the embeddings of its title or abstract.
In particular, the input to the BERT is $v$'s title or abstract text which is represented by a sequence of terms, i.e., $\mathbf{W} = [\mathbf{w}_1, \mathbf{w}_2,\cdots, \mathbf{w}_{n_w}]$, where $\mathbf{w}_i$ is the $i$-th term and $n_w$ is the length of text.
Then, two dense embedding layers will transform the BERT's outputs into $\mathbf{H} = [\mathbf{h}_1, \mathbf{h}_2, ... , \mathbf{h}_{n_w}]$:
\begin{equation}
\mathbf{h}_i = \mathbf{V}_1^n \times {\rm LeakyReLU}(\mathbf{V}_2^n \times {\rm BERT}(\mathbf{w}_i) + \mathbf{b}_2^n) + \mathbf{b}_1^n,
\end{equation}
where $\mathbf{V}_1^n$, $\mathbf{V}_2^n$, $\mathbf{b}_2^n$, and $\mathbf{b}_1^n$ are learnable model parameters.

To further extract the interactive semantics between terms, multi-head self-attention is used to convert $\mathbf{H}$ into $\mathbf{T}$:
\begin{equation}
\mathbf{T} = {\rm MultiHeadAttention}(\mathbf{H}),
\end{equation}
where $\mathbf{T} = [\mathbf{t}_1, \mathbf{t}_2, \cdots, \mathbf{t}_{n_w}]$.

As each entity $\mathbf{e}_i$ ($i=1,2,\cdots,n_e$) in persona $\mathbf{E}_u$ may yield valuable additional information to complement the surface text $\mathbf{w}_j$ ($j=1,2,\cdots,n_w$), we can use repeated attention for each ($\mathbf{e'}_i$, $\mathbf{W}$) pair to strengthen $\mathbf{T}$ into $\mathbf{R}$:
\begin{equation}
\mathbf{e'}_i = {\rm LeakyReLU}(\mathbf{V}_3^n \times \mathbf{e}_i + \mathbf{b}_3^n),
\end{equation}
\begin{equation}
\widetilde{\alpha}_{ij}^n = \frac{\exp(\mathbf{e'}_i^T \mathbf{Q}^n \mathbf{t}_j)}{\sum_{k=1}^{n_w} \exp(\mathbf{e'}_i^T \mathbf{Q}^n \mathbf{t}_k)},
\end{equation}
\begin{equation}
\mathbf{r}_i = \sum_{j=1}^{n_w} \widetilde{\alpha}_{ij}^n \mathbf{t}_j,
\end{equation}
where $\mathbf{R}=\{{\mathbf{r}_1,\mathbf{r}_2,\cdots,\mathbf{r}_{n_e}}\}$ is the richer and more personalized representation for the textual content $\mathbf{W}$, $\widetilde{\alpha}_{ij}^n$ is the attention weight between the $i$-th entity and the $j$-th term, and $\mathbf{V}_3^n$, $\mathbf{b}_3^n$, $\mathbf{Q}^n$ are learnable model parameters.

To generate the final news item representation $\mathbf{r}$, we use another attention layer to summarize the above information:
\begin{equation}
\beta_i^n = \mathbf{q}_n^T \times {\rm tanh}(\mathbf{V}_4^n \times \mathbf{r}_i + \mathbf{b}_4^n), 
\end{equation}
\begin{equation}
\widetilde{\beta}_i^n = \frac{\exp(\beta_i^n)}{\sum_{j=1}^{n_e} \exp(\beta_j^n)},
\end{equation}
\begin{equation}
\mathbf{r} = \sum_{i=1}^{n_e} \widetilde{\beta}_i^n \mathbf{r}_i,
\end{equation}
where $\widetilde{\beta}_i^n$ is the attention weight of the $i$-th representation, and $\mathbf{q}_n^T$, $\mathbf{V}_4^n$, $\mathbf{b}_4^n$ are learnable model parameters.

\subsection{User Encoder}
The user encoder is designed to model a user's reading interests with his/her explicit persona as well as his/her top-$n_u$ (usually $n_u=20$) recently-read news items. 
The corresponding sub-network module is shown on the right side of \cref{fig:PerCoNet}.
Here a user's persona can reflect that user's \emph{long-term} stable preferences over news items~\cite{Mingxiao-LSTUR-ACL-2019}.
By combing users' explicit persona with their recently-read news items, the user encoder can dynamically characterize both long-term and short-term user interests for better personalization.  

Given a user's explicit persona $\mathbf{E}_u$ and his/her recently-read news items $V=\{v_1,v_2,\cdots,v_{n_u}\}$, we supply them to the news encoder and use a multi-head self-attention layer to generate an enhanced semantic representation $\mathbf{Z}$:
\begin{equation}
\mathbf{Z} = \rm{MultiHeadAttention}(\rm{NewsEncoder}(\mathbf{V}, \mathbf{E}_u)),
\end{equation} 
where $\mathbf{Z}=[\mathbf{z}_1,\mathbf{z}_2,\cdots,\mathbf{z}_{n_u}]$.

Next, similar to the news encoder, each entity $\mathbf{e}_i$ can attend over the enriched embeddings $\mathbf{Z}$ and output a personalized representation: 
\begin{equation}
\alpha_{ij}^u = \mathbf{q}_1^T \times {\rm LeakyReLU}(\mathbf{V}_1^u \times (\mathbf{e}_i \oplus \mathbf{z}_j) + \mathbf{b}_1^u),
\end{equation} 
\begin{equation}
\widetilde{\alpha}_{ij}^u = \frac{\exp(\alpha_{ij}^u)}{\sum_{k=1}^{n_u} \exp(\alpha_{ik}^u)},
\end{equation}
\begin{equation}
\mathbf{o}_i = \sum_{j=1}^{n_u} {\widetilde{\alpha}_{ij}^u \mathbf{z}_j},
\end{equation}
where $\widetilde{\alpha}_{ij}^u$ is the attention weight between the $i$-th entity and the $j$-th news representation, and $\mathbf{q}_1$, $\mathbf{V}_1^u$, $\mathbf{b}_1^u$ are learnable model parameters.
Furthermore, let $\mathbf{O} = [\mathbf{o}_1, \mathbf{o}_2, \cdots, \mathbf{o}_{n_e}]$.

To generate the final user interest representation $\mathbf{u}$, we use another attention layer to further condense the above information:
\begin{equation}
\beta_i^u = \mathbf{q}_2^T \times {\rm{tanh}}(\mathbf{V}_2^u \times \mathbf{o}_i + \mathbf{b}_2^u),
\end{equation}
\begin{equation}
\widetilde{\beta}_i^u = \frac{{\rm{exp}}(\beta_i^u)}{\sum_{j=1}^{n_e} {\rm{exp}}(\beta_j^u)},
\end{equation}
\begin{equation}
\mathbf{u} = \sum_{i=1}^{n_e} \widetilde{\beta}_i^u \mathbf{o}_i,
\end{equation}
where $\widetilde{\beta}_i^u$ is the attention weight of the $i$-th user representation, and $\mathbf{q}_2$, $\mathbf{V}_2^u$, $\mathbf{b}_2^u$ are learnable model parameters.

\subsection{Click Probability Prediction}

The general framework shown in the middle of \cref{fig:PerCoNet} shows how the click probability is calculated from the corresponding news item, the user's recently-read news items, and his/her explicit persona.

Specifically, for a candidate news item, its title $\mathbf{T}_c$ and the user's persona $\mathbf{E}_u$ are fed into the news encoder to get the news representation $\mathbf{r}_c$. 
Meanwhile, the titles $\{\mathbf{T}_1,\mathbf{T}_2,\cdots,\mathbf{T}_{n_u}\}$ of the user's recently-read news items and his/her persona are fed into the user encoder to get the user representation $\mathbf{u}$. 
Then we concatenate the news and user representations together, and enter them into a multi-layer perception (MLP) layer to calculate the click probability $\widehat{y}$, i.e., how likely the user is going to click on that particular news item:
\begin{equation}
\widehat{y} = {\rm sigmoid}(\mathbf{q}_c^T \times {\rm LeakyReLU} (\mathbf{V}^c \times (\mathbf{u} \oplus \mathbf{r}^c) + \mathbf{b}^c)),
\end{equation}
where $\mathbf{q}_c$, $\mathbf{V}^c$, $\mathbf{b}^c$ are learnable parameters.

\subsection{Cross-View Contrastive Learning}
\paragraph{Remarks.} 
The simplest way to combine `\texttt{title}' and `\texttt{abstract}' of a news item is to just concatenate them together as the input of news encoder, as in most existing work~\cite{Huifeng-DeepFM-IJCAI-2017,Chuhan-NAML-IJCAI-2019,Xuanyu-EEG-WWW-2021,Rongyao-MINS-ICASSP-2022,Qiwei-MTRec-ACL-2022}. 
Here we choose not to adopt that method of direct combination, but to treat `\texttt{title}' and `\texttt{abstract}' as two distinct though semantically-related fields which motivates us to design a new cross-view contrastive learning task for better news/user encodings.
A novel cross-view contrastive learning sub-network has been designed and incorporated into the learning framework as shown in the middle of \cref{fig:PerCoNet}.  
It is inspired by the observation that every news item has two separate but correlated views: `\texttt{title}' and `\texttt{abstract}'. 
Therefore, when the user encoder is summarizing a user's recently-read news items, if those articles' titles are replaced with their corresponding abstracts, then the outputs should be similar within the same user but dissimilar between different users. 

Specifically, for a user $\mathbf{u}$, assume that his/her recently-read news items' titles and abstracts are $\mathbf{T}=\{\mathbf{T}_1,\mathbf{T}_2,\cdots,\mathbf{T}_{n_u}\}$ and $\mathbf{A}=\{\mathbf{A}_1,\mathbf{A}_2,\cdots,\mathbf{A}_{n_u}\}$ respectively. 
First, we perform dropout over the entire set of titles $\mathbf{T}$ and shuffle the remaining ones to get a subset $\mathbf{T}_{sub}$.
Then, we can make $\mathbf{u}^a$ and $\mathbf{u}^t$ through the following operations:
\begin{equation}
\mathbf{u}^a = {\rm UserEncoder}(\mathbf{A},\mathbf{E}_u),
\end{equation}
and
\begin{equation}
\mathbf{u}^t = {\rm UserEncoder}(\mathbf{T}_{sub},\mathbf{E}_u).
\end{equation}

Following the user encoder, an MLP layer will be able to convert $\mathbf{u}^a$ and $\mathbf{u}^t$ into $\mathbf{u}_l$ and $\mathbf{u}_l^+$ respectively:
\begin{equation}
\mathbf{u}_l = {\rm LeakyReLU}(\mathbf{V}_1^l \times {\rm LeakyReLU}(\mathbf{V}_2^l \times \mathbf{u}^a + \mathbf{b}_2^l) + \mathbf{b}_1^l),
\end{equation}
and
\begin{equation}
\mathbf{u}_l^+ = {\rm LeakyReLU}(\mathbf{V}_1^l \times {\rm LeakyReLU}(\mathbf{V}_2^l \times \mathbf{u}^t + \mathbf{b}_2^l) + \mathbf{b}_1^l),
\end{equation}
where $\mathbf{V}_1^l$, $\mathbf{V}_2^l$, $\mathbf{b}_1^l$, $\mathbf{b}_2^l$ are learnable model parameters.

Obviously, for the same user $\mathbf{u}$, the output $\mathbf{u}_l^+$ can be treated as the cross-view positive example of the output $\mathbf{u}_l$. 
Moreover, there are a lot of other users $\mathbf{u}_i$, so many cross-view negative examples $\mathbf{u}_i^-$ could be sampled for contrastive learning formulated as follows:
\begin{equation}
Loss_{CL} = -\log\frac{\exp\left(\frac{\mathbf{u}_l^T \mathbf{u}_l^+}{\tau}\right)}{\exp\left(\frac{\mathbf{u}_l^T \mathbf{u}_l^+}{\tau}\right) + {\sum_{i=1}^{n_b}\exp\left(\frac{\mathbf{u}_l^T \mathbf{u}_i^-}{\tau}\right)}},
\end{equation}
where $\tau$ is a temperature hyperparameter, and $n_b$ denotes the number of cross-view negative samples.

\subsection{Model Training}

For the news recommendation task, we use the following negative sampling strategy for model training. 
Specifically, given a user, we regard the clicked news items as positive examples and the other news items as negative examples.
For each positive example $v_i$ in a mini-batch $S$, we randomly sample $H$ negative examples.
Thus the contrastive loss can be written as:
\begin{equation}
Loss_{REC} = -\sum_{i=1}^{\left\vert S \right\vert} \log\frac{\widehat{y_i}}{\widehat{y_i} + \sum_{j=1}^H \widehat{y_{j}}},
\end{equation}
where $|S|$ is the size of the training set $S$, and $H$ is the number of negative examples for each positive example.

Since the cross-view contrastive learning task is based on the shared news and user encoders, it could be jointly learned along with the news recommendation task.
In the end, the overall loss of our PerCoNet model is:
\begin{equation}
Loss = Loss_{REC} + \lambda \times Loss_{CL},
\end{equation}
where $\lambda$ is a hyper-parameter, $Loss_{REC}$ is the loss for the main task of news recommendation and $Loss_{CL}$ is the loss for the auxiliary task of cross-view contrastive learning.

\section{Experiments}
\label{sec:Experiments}

In this section, we testify PerCoNet's high effectiveness by comparing it with several competitive methods. 
All the Python source \textbf{code} and the processed \textbf{datasets} will be made publicly available online for the reproducibility of our experiments.

\subsection{Datasets}

The \textbf{MIND}\footnote{\url{https://msnews.github.io/}} dataset~\cite{Fangzhao-MIND-ACL-2020}
contains one million users and 160k+ English news articles from Microsoft News (in the period from 10/12/2019 to 11/22/2019). 
Each news article has rich textual attributes including title, abstract, and entities. 
For easy comparison with other existing methods, we evaluate our proposed PerCoNet model on MIND-small~\cite{Qi-UNBERT-IJCAI-2021} 
which is a subset of MIND constructed by randomly sampling $50,000$ users and their daily behaviors.
The samples in the first five weeks are used for training, and those in the last week are reserved for testing. 

The \textbf{Adressa}\footnote{\url{http://reclab.idi.ntnu.no/dataset/}} dataset~\cite{Jon-Adressa-WI-2017} 
contains a large number of Norwegian news articles and many anonymous users.
There exist two versions of Adressa~\cite{Yong-HyperNews-IJCAI-2020}: Adressa-1week and Adressa-4week. 
We conduct experiments on the former dataset, in which the first six days' events are used for training and the last day's events for testing.

\cref{tab:datasets-statistics} summarizes the statistics of these two popular datasets for personalized news recommendation.

\begin{table}[!tb]
	\centering%
	\renewcommand\arraystretch{1.0}
	\setlength{\tabcolsep}{6mm}{
	\begin{tabular}{lll|rl|rl|}
		\toprule
		\multicolumn{3}{|l|}{Statistics}                   & \multicolumn{2}{|c|}{\textbf{MIND}} & \multicolumn{2}{|c|}{\textbf{Adressa}} \\ 
		\midrule
		\multicolumn{3}{|l|}{\#users}                            & \multicolumn{2}{r|}{50,000}        & \multicolumn{2}{r|}{601,215}          \\
		\multicolumn{3}{|l|}{\#news}                            & \multicolumn{2}{r|}{65,238}        & \multicolumn{2}{r|}{17,692}           \\
		\multicolumn{3}{|l|}{\#entities-per-news} & \multicolumn{2}{r|}{3.04}          & \multicolumn{2}{r|}{7.42}             \\
		\multicolumn{3}{|l|}{\#words-per-news-title}          & \multicolumn{2}{r|}{11.52}         & \multicolumn{2}{r|}{6.63}             \\
		\multicolumn{3}{|l|}{\#click-behaviors}                          & \multicolumn{2}{r|}{347,727}       & \multicolumn{2}{r|}{3,123,261}        \\ 
		\bottomrule
	\end{tabular}}
	\caption{The statistics of the datasets for our experiments: `\#users' is the total number of users; `\#news' is the total number of news; `\#entities-per-news' is the average number of entities in each news; `\#words-per-news-title' is the average number of words in each news title; and `\#click-behaviors' is the total number of all news-clicking behaviors.}\label{tab:datasets-statistics}
\end{table}

\begin{table*}[!tb]
	\centering
	\renewcommand\arraystretch{1.0}
	\setlength{\tabcolsep}{4.4mm}{
		\begin{tabular}{|ll|cccc|cccc|}
			\toprule
			\multicolumn{2}{|c|}{\multirow{2}{*}{Method}} & \multicolumn{4}{c|}{\textbf{MIND}}                                         & \multicolumn{4}{c|}{\textbf{Adressa}}                                       \\ \cline{3-10} 
			\multicolumn{2}{|l|}{}                        & AUC            & MRR            & nDCG@5         & nDCG@10        & AUC            & MRR            & nDCG@3         & nDCG@5         \\ 
			\midrule
			\multicolumn{2}{|l|}{LibFM~\cite{Steffen-LibFM-TIST-2012}}                   & 57.93          & 24.64          & 26.53          & 33.35          & 55.98          & 49.73          & 48.78          & 62.12          \\
			\multicolumn{2}{|l|}{DeepFM~\cite{Huifeng-DeepFM-IJCAI-2017}}                  & 58.18          & 25.87          & 27.57          & 34.25          & 59.95          & 59.27          & 55.06          & 69.18          \\ 
			\multicolumn{2}{|l|}{DKN~\cite{Hongwei-DKN-WWW-2018}}                     & 62.14          & 27.83          & 30.04          & 36.70          & 65.49          & 63.77          & 60.71          & 72.60          \\
			\multicolumn{2}{|l|}{LSTUR~\cite{Mingxiao-LSTUR-ACL-2019}}                   & 64.83          & 30.41          & 33.37          & 39.81          & 66.17          & 64.27          & 62.15          & 72.86          \\
			\multicolumn{2}{|l|}{NAML~\cite{Chuhan-NAML-IJCAI-2019}}                    & 66.02          & 32.13          & 35.41          & 41.37          & 67.42          & 62.97          & 62.30          & 72.10          \\
			\multicolumn{2}{|l|}{NRMS~\cite{Chuhan-NRMS-EMNLP-2019}}                    & 65.56          & 30.48          & 33.45          & 39.97          & 68.24          & 66.00          & 65.31          & 74.80          \\
			\multicolumn{2}{|l|}{KRED~\cite{Danyang-KRED-RecSys-2020}}                    & 66.27          & 32.27          & 35.53          & 41.52          & 69.76          & 66.01          & 66.11          & 74.82          \\
			\multicolumn{2}{|l|}{EEG~\cite{Xuanyu-EEG-WWW-2021}}                  & 66.47          & 31.80          & 34.85          & 41.23          & 68.49          & 65.71          & 63.69          & 73.67          \\
			\multicolumn{2}{|l|}{UNBERT~\cite{Qi-UNBERT-IJCAI-2021}}                  & 67.57          & 32.40          & 35.85          & 42.26          & 74.47          & 68.23          & 68.46          & 74.82          \\ 
			\bottomrule
			\multicolumn{2}{|l|}{\textbf{PerCoNet$^*$}}                      & \textbf{68.93} & \textbf{33.40} & \textbf{36.93} & \textbf{43.28} & \textbf{77.26} & \textbf{72.20} & \textbf{70.18} & \textbf{76.22} \\ 
			\bottomrule
	\end{tabular}}
	\caption{The news recommendation performances of different methods on MIND and Adressa. Note that boldface indicates the best results, and $^*$ denotes that PerCoNet's performance improvement over each and every baseline method is statistically significant with $p<0.05$.}\label{tab:recommendation-competition}
\end{table*}

\subsection{Competitors and Metrics}
We compare our PerCoNet with several state-of-the-art methods including:
\textbf{LibFM\footnote{\url{https://github.com/srendle/libfm}}}~\cite{Steffen-LibFM-TIST-2012},
\textbf{DeepFM\footnote{\url{https://github.com/ChenglongChen/tensorflow-DeepFM}}}~\cite{Huifeng-DeepFM-IJCAI-2017}, 
\textbf{DKN\footnote{\url{https://github.com/hwwang55/DKN}}}~\cite{Hongwei-DKN-WWW-2018},
\textbf{LSTUR\footnote{\url{https://github.com/microsoft/recommenders}\label{attentive-news-recommendation}}}~\cite{Mingxiao-LSTUR-ACL-2019},
\textbf{NAML\textsuperscript{\ref{attentive-news-recommendation}}}~\cite{Chuhan-NAML-IJCAI-2019},
\textbf{NRMS\textsuperscript{\ref{attentive-news-recommendation}}}~\cite{Chuhan-NRMS-EMNLP-2019},
\textbf{KRED\footnote{\url{https://github.com/danyang-liu/KRED}}}~\cite{Danyang-KRED-RecSys-2020},
\textbf{UNBERT\footnote{\url{https://github.com/huawei-noah/benchmark/tree/main/FuxiCTR/model_zoo/UNBERT}}}~\cite{Qi-UNBERT-IJCAI-2021}, and
\textbf{EEG\footnote{We implement it with pytorch code.}}~\cite{Xuanyu-EEG-WWW-2021}.

The metrics used in our experiments are nDCG@top$N$ (with top$N$=5,10 for MIND and top$N$=3,5 for Adressa), AUC, and MRR, all of which are commonly used metrics for recommendation performance evaluation. 
Each experiment was repeated $10$ times and the final score is their average.

\subsection{Settings}
In our experiments, the batch size and the negative sampling ratio are set to 64 and 4 respectively. 
The dropout probability is $0.2$.
The optimizer is Adam~\cite{Diederik-Adam-ICLR-2015}, and the learning rate is $8e-5$.
The maximum length of a news title or abstract is $20$.
The maximum number of a user's clicked news is $20$. 
The pre-trained BERT model is \textit{roberta-base\footnote{\url{https://huggingface.co/models}\label{bert-model}}}~\cite{Yinhan-RoBERTa-CoRR-2019} for MIND, and \textit{bert-base-multilingual-cased\textsuperscript{\ref{bert-model}}}~\cite{Jacob-BERT-NAACL-2019} for Adressa.
In light of BERTs, we only fine-tune the last three Transformer~\cite{Ashish-AttentionAllNeed-NeurIPS-2017} layers. 
The hyperparameters $\lambda$ and $\tau$ are tuned to 1 and 0.05, respectively.
Please refer to our code for more experiment configuration details.

For the selected competitors, we adopt the best configurations as described in each method's corresponding paper.

\subsection{Results}

\cref{tab:recommendation-competition} shows the performance scores of the selected methods on MIND and Adressa datasets, from which we have the following observations.

First, the methods based on deep neural networks (e.g., DKN, LSTUR, NAML and UNBERT) perform much better than the traditional matrix factorization methods replying on handcrafted features (such as LibFM), which suggests that neural networks are more capable to model complex news datasets than the traditional shallow methods requiring manual feature engineering.

Second, UNBERT, which introduces the pre-trained BERT model, beats other baseline methods that initialize word representations by Glove vectors (such as NAML and NRMS). 
This is probably owing to that the pre-trained model can enhance the textual representations with external rich language knowledge.

Next, it's worth making a comparison between the very recently proposed EEG and our PerCoNet. They both utilize explicit and implicit text information (such as `\texttt{title}', `\texttt{abstract}', and `\texttt{entity}') but in different ways. With respect to EEG, it fuses `\texttt{title}', `\texttt{abstract}', and also `\texttt{body}' via CNNs and attention modules for news encoder, and constructs entity graph for user interests; while in light of PerCoNet, it enhances implicit semantics via cross-view contrastive learning between `\texttt{title}' and `\texttt{abstract}', and builds persona with explicit entities to learn more personalized representations for both news and user encoders.
The significant advantages of PerCoNet over EEG tell us that the organization and utilization of limited data sources really play a critical role in news recommendation. 

Finally, our PerCoNet method achieves the best performances w.r.t. several metrics on all these datasets, which verifies its effectiveness for personalized news recommendation.
Such performance gains are attributed to the persona-aware news and user encoders which better model news content and user interests, and also the cross-view contrastive learning task  which is learned along with the main click probability prediction task in a mutual-reinforcing multi-task learning manner.

\begin{figure*}[!tb]
	\centering
	\subfloat[$\lambda$ (MIND)]{\includegraphics[width=0.66\columnwidth]{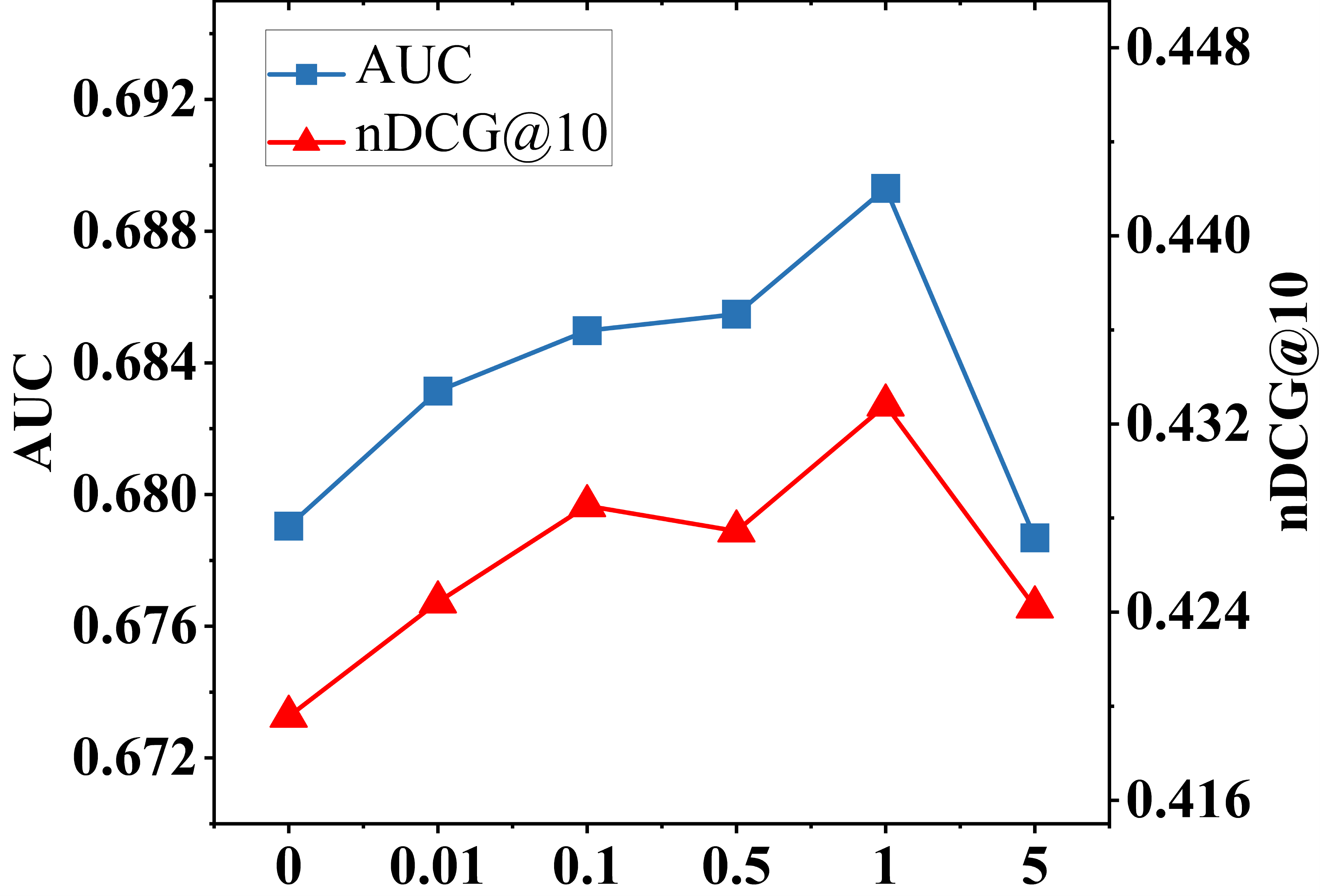}\label{fig:hyperparameter-lambda}}\hspace{4mm}
	\subfloat[top-$K$ (MIND)]{\includegraphics[width=0.66\columnwidth]{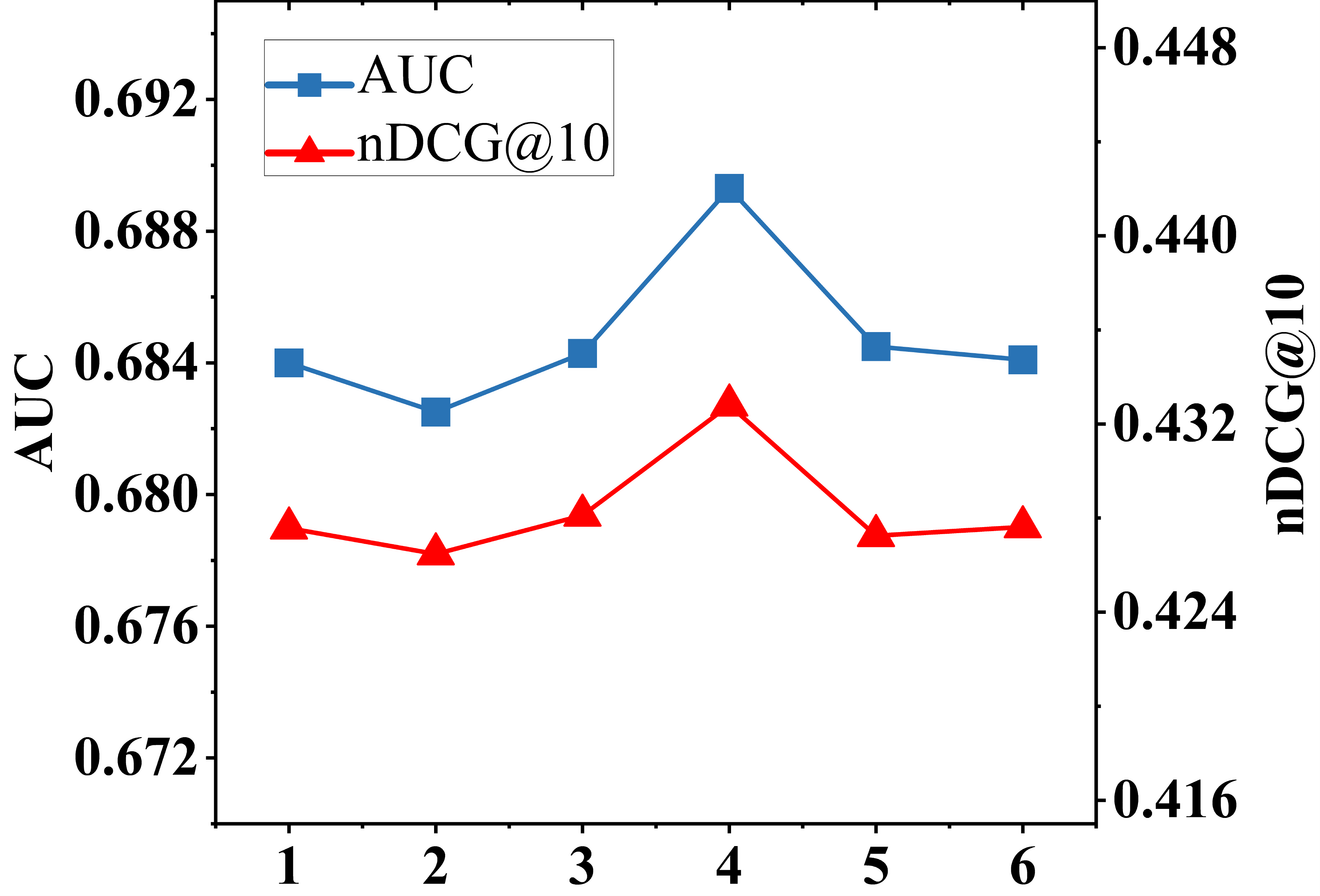}\label{fig:hyperparameter-top-K}}\hspace{4mm}
	\subfloat[top-$G$ (MIND)]{\includegraphics[width=0.66\columnwidth]{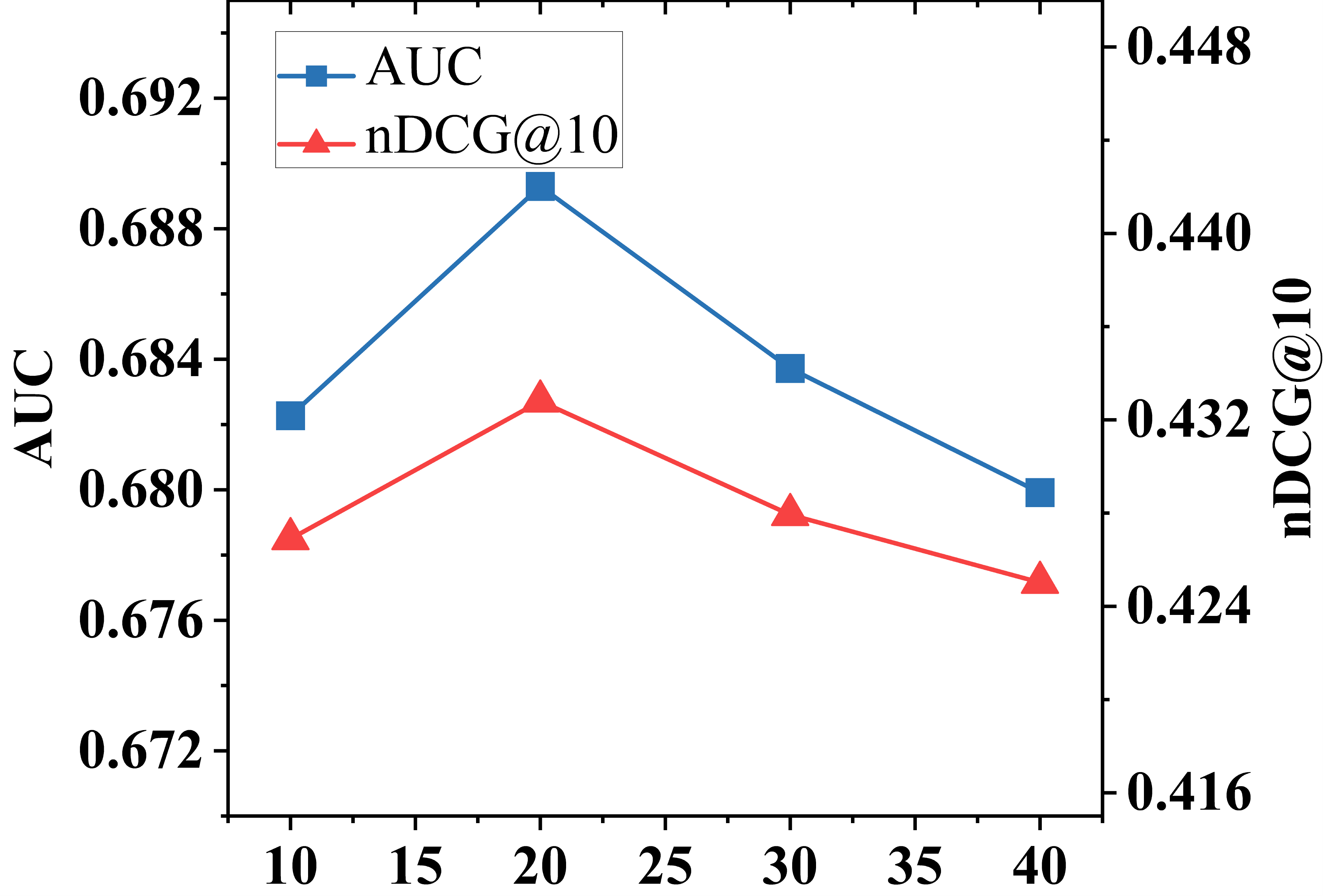}\label{fig:hyperparameter-top-G}}
	
	\subfloat[$\lambda$ (Adressa)]{\includegraphics[width=0.66\columnwidth]{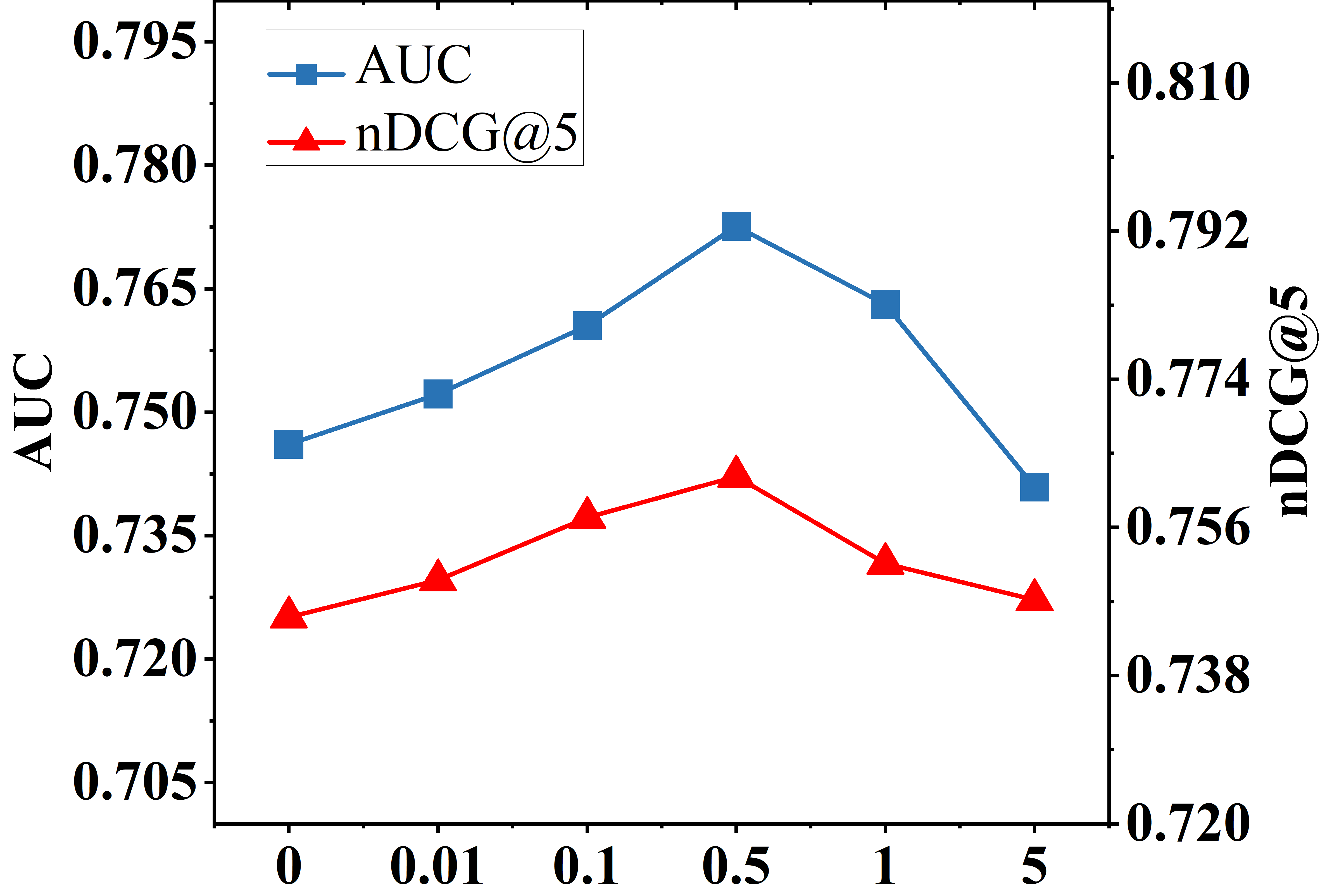}\label{fig:adressa_hyperparameter-lambda}}\hspace{4mm}
	\subfloat[top-$K$  (Adressa)]{\includegraphics[width=0.66\columnwidth]{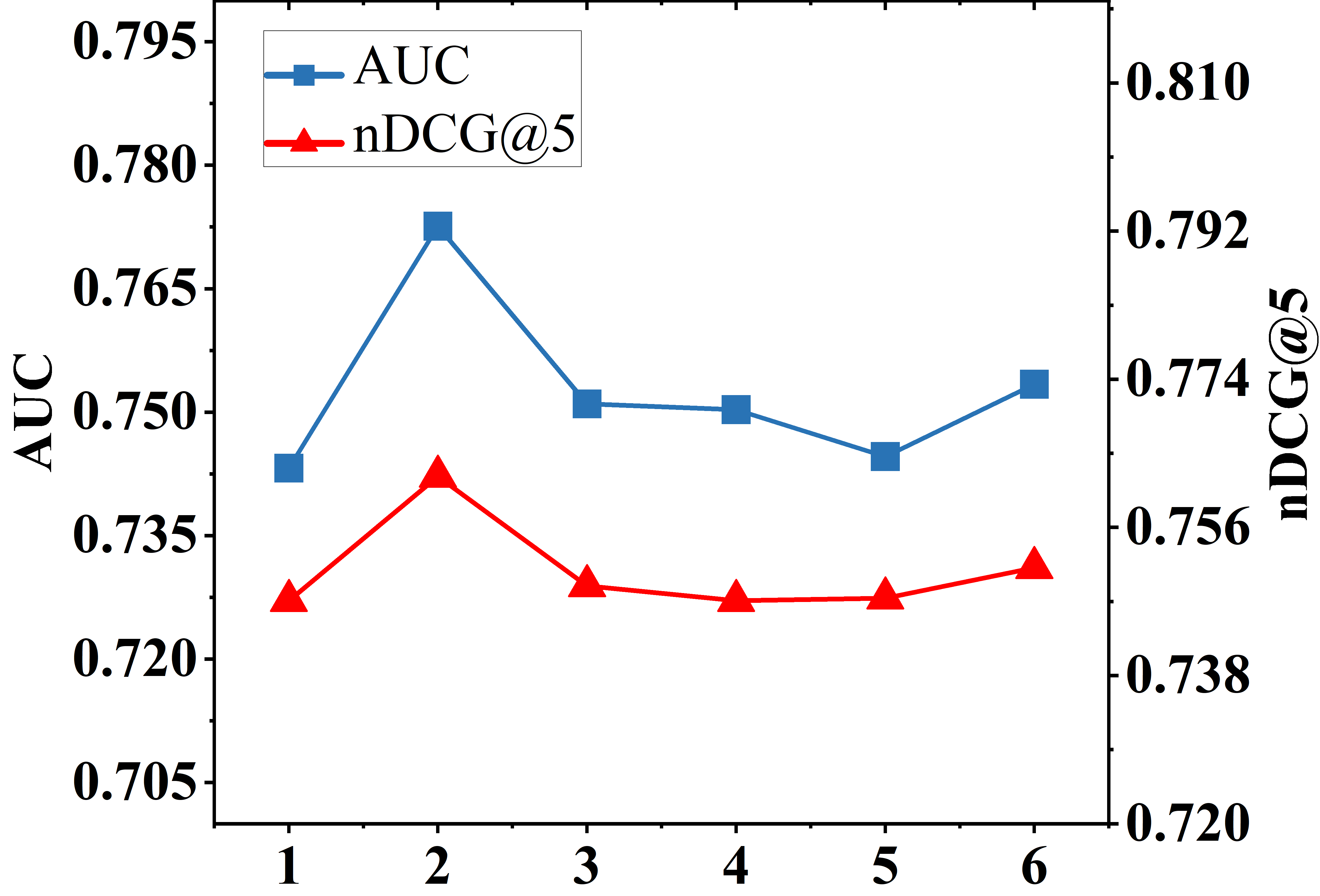}\label{fig:adressa_hyperparameter-top-K}}\hspace{4mm}
	\subfloat[top-$G$  (Adressa)]{\includegraphics[width=0.66\columnwidth]{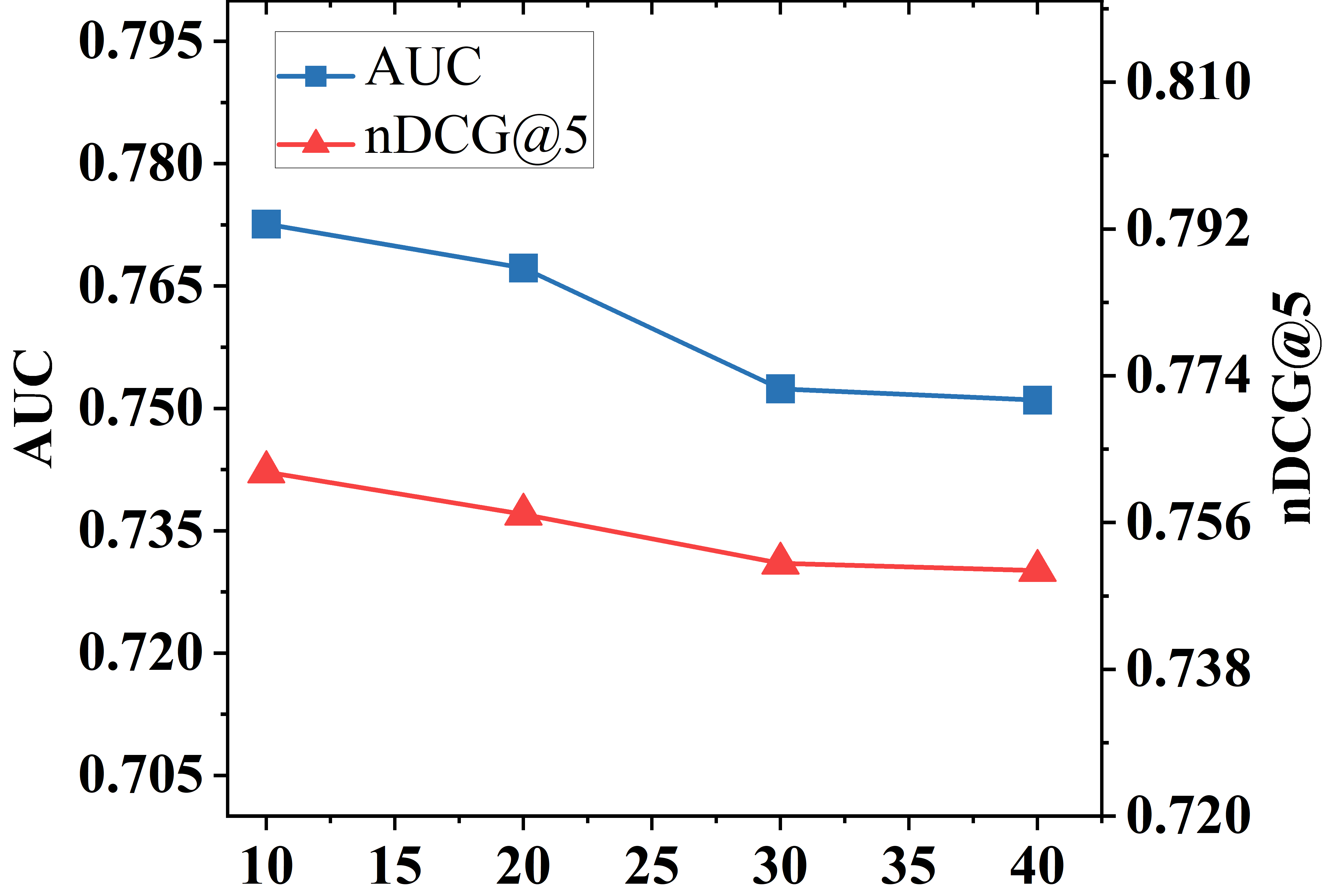}\label{fig:adressa_hyperparameter-top-G}}
	\caption{How the performance of PerCoNet is affected by the hyperparameters $\lambda$, top-$K$ and top-$G$.}
	\label{fig:parameter-sensitivity-analysis}
\end{figure*}

\subsection{Hyperparameter Analysis}

Here, we do experiments to study the effects of PerCoNet's three most important hyperparameters $\lambda$, top-$K$ and top-$G$.

First, $\lambda$ is the balance weight that controls the proportion of multi-task loss.
The AUC scores of our PerCoNet with different $\lambda$ values on the MIND dataset is shown in \cref{fig:hyperparameter-lambda}.
It can be seen that  $\lambda = 1$ yields competitive results. 

Second, we analyze the influence of the number (i.e., top-$K$) of entities from each news item on the building of personae.
As can be seen from \cref{fig:hyperparameter-top-K}, taking top-$4$ entities from each news item would be optimal.

Next, we examine the number (i.e., top-$G$) of news items from which we extract entities for personae construction.
As shown in \cref{fig:hyperparameter-top-G}, when the number of news items is too small, the performance of PerCoNet is not good due to the lack of user information;
whereas when the number of news is too large, personae probably contains some noisy entities and thus the performance also falls.

Besides, the hyperparameter analysis on the Adressa dataset is similar as that on the Mind dataset.

In the end, we set $\lambda$=1, top-$K$=4, and top-$G$=20 on the MIND dataset, and $\lambda$=0.5, top-$K$=2, and top-$G$=10 on the Adressa dataset for competitive results.

\subsection{Ablation Study}
To study the specific effectiveness of the persona-aware module and the contrastive learning module, we conduct a series of ablation experiments:
\begin{enumerate}[(i)]
	\item PerCoNet without persona component (PerCoNet$_{{no-persona}}$),
	
	\item PerCoNet without cross-view contrastive learning between `\texttt{title}' and `\texttt{abstract}' (PerCoNet$_{{no-CL}}$),
	
	\item PerCoNet without these two components (PerCoNet$_{{no-both}}$),
\end{enumerate}  
on the two real-world news datasets.

\begin{figure*}[!tb]
	\centering
	\subfloat[Comparing the complete version of PerCoNet and its incomplete variant with no explicit persona, or with no contrastive learning, or with neither.]{\includegraphics[width=1.0\columnwidth]{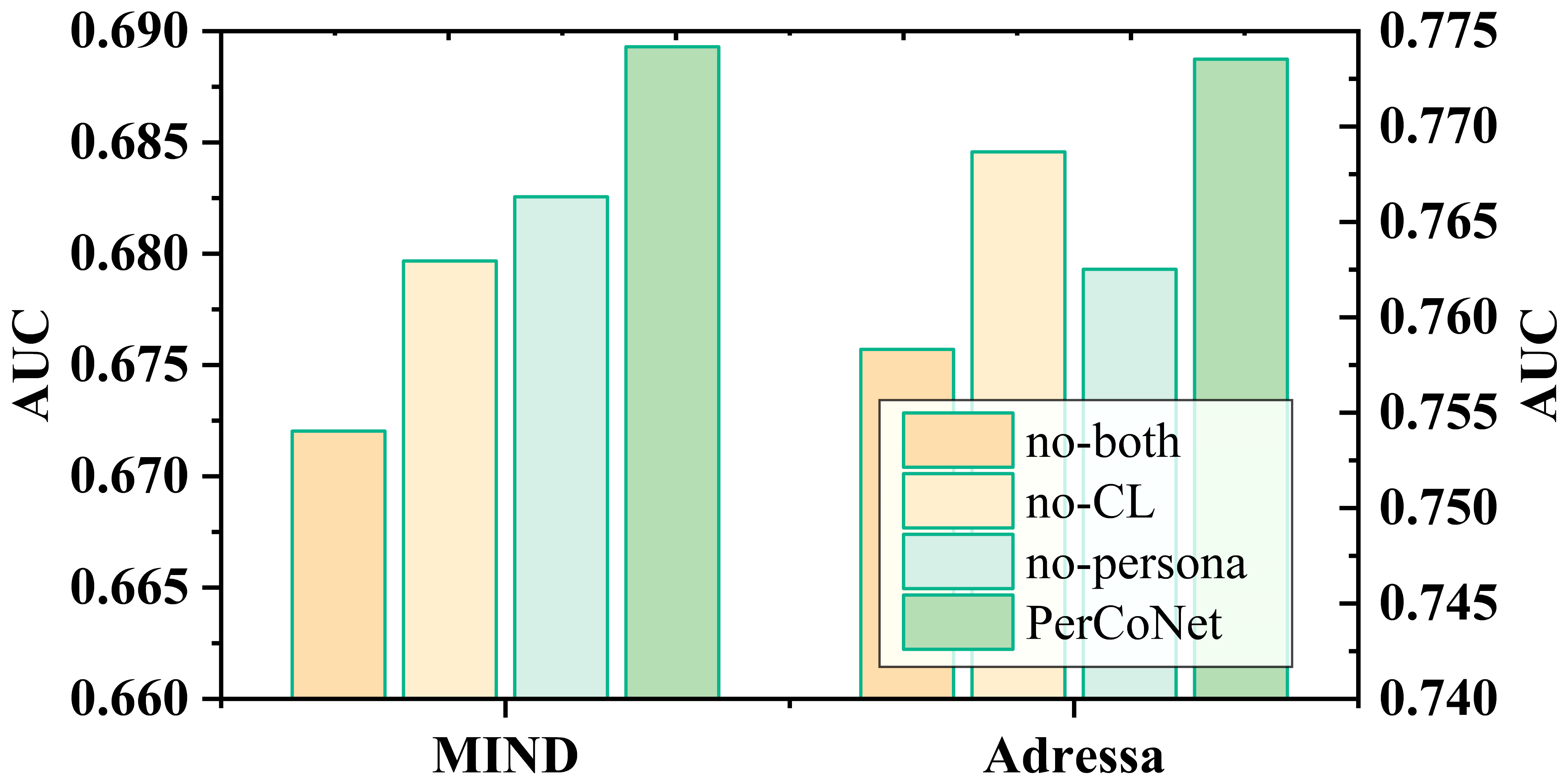}\label{fig:ablation-studies}}\hspace{6mm}
	\subfloat[Which way is better for the effective utilization of both `\texttt{title}' and `\texttt{abstract}' information: contrastive learning, or direct combination?]{\includegraphics[width=1.0\columnwidth]{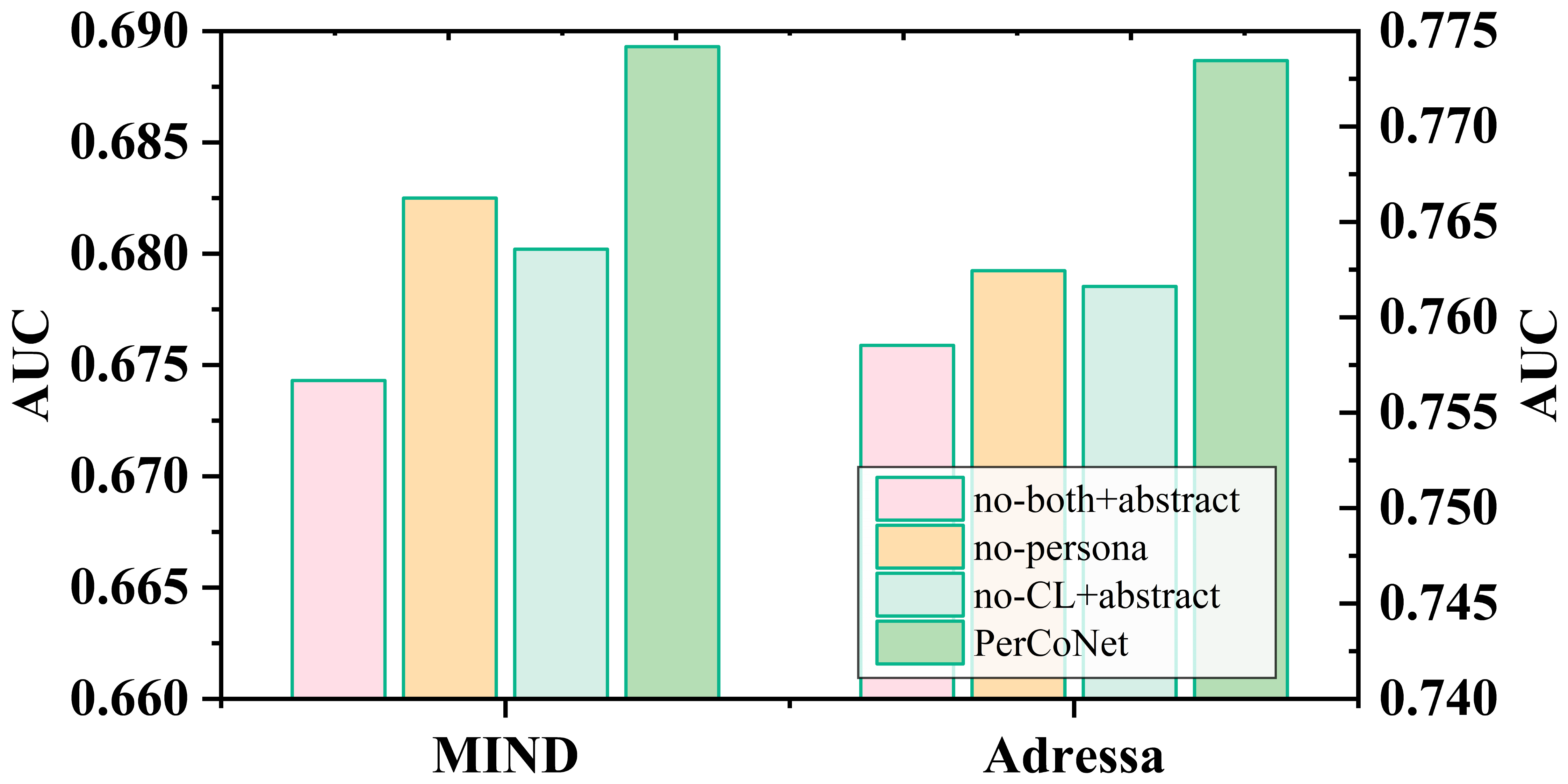}\label{fig:ablation-studies_CL_}}
	\caption{The ablation studies of PerCoNet.}
	\label{fig:ablation}
\end{figure*}

Their results are plotted in \cref{fig:ablation-studies}.
First, ${\rm PerCoNet}_{{\rm no-CL}}$ outperforms ${\rm PerCoNet}_{{\rm no-both}}$, which indicates the effectiveness of the persona-aware network for news and user representations.
Second, ${\rm PerCoNet}_{{\rm no-persona}}$ performs better than ${\rm PerCoNet}_{{\rm no-both}}$, which tells us that the cross-view contrastive learning module can indeed improve news and user representations via joint learning.
Finally, the complete version of PerCoNet achieves the best performance on both datasets, which proves that the explicit persona-aware module and cross-view contrastive learning module together could further boost the news recommendation performance.

Moreover, we would like to find out which is the best way to make use of the `\texttt{title}' and `\texttt{abstract}' information, just combining them directly, or more sophisticated cross-view contrastive learning?
To answer this question, we design two more experiments:
\begin{enumerate}[(i)]
	\item PerCoNet v.s. PerCoNet$_{no-CL+abstract}$,
	\item PerCoNet$_{no-persona}$ v.s. PerCoNet$_{no-both+abstract}$,
\end{enumerate} 
on these two datasets. 

To be more specific, here combining `\texttt{title}' and `\texttt{abstract}' directly means that we treat `abstact' as `\texttt{title}' in the news encoder (see the left corner of \cref{fig:PerCoNet}) without extra cross-view contrastive learning between them; and `the more sophisticated cross-view contrastive learning' corresponds to that we only use the `\texttt{title}' field in the news encoder and exploit extra cross-view contrastive learning between `\texttt{title}' and `\texttt{abstract}' as an auxiliary task (see the middle part of \cref{fig:PerCoNet}). Therefore, the above two comparative experiments examine how to effectively use the `\texttt{abstract}' field under two different base models (i.e., PerCoNet and PerCoNet$_{no-persona}$).

Their results are plotted in \cref{fig:ablation-studies_CL_}.
Obviously, no matter it's on MIND or Adressa, PerCoNet (PerCoNet$_{no-persona}$) performs better than PerCoNet$_{no-CL+abstract}$ (PerCoNet$_{no-both+abstract}$), which tells us that exploiting the `\texttt{title}' and `\texttt{abstract}' of each news item via cross-view contrastive learning works better than using a straightforward combination of them.

%

\begin{figure*}[!tb]
	\centering
	\includegraphics[width=0.99\textwidth]{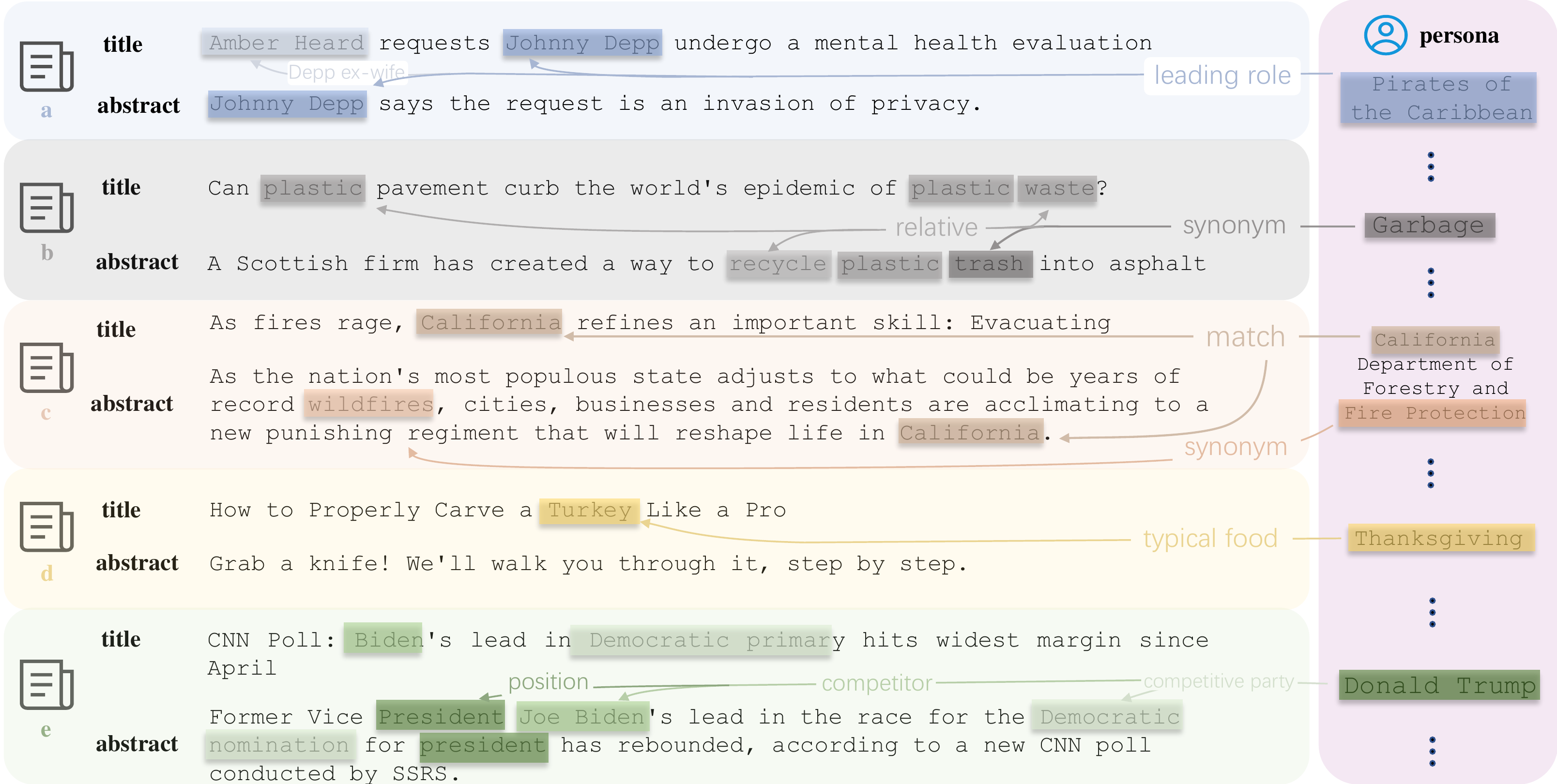}
	\caption{The news items (left) recommended by PerCoNet for a particular user based on his/her explicit persona (right).}\label{fig:Person_Recomm}
\end{figure*}

\subsection{A Real Case}

\cref{fig:Person_Recomm} shows how each of the news items recommended by PerCoNet for a particular user (randomly picked from the aforementioned real-world dataset MIND) are connected to the entities in his/her explicit persona.
For example, news item \texttt{(a)}	is chosen by PerCoNet to be recommended to the user probably because it mentions the actor ``Johnny Depp'' who played the leading role in the movie ``Pirates of the Caribbean'' which the user likes.
It can be observed that PerCoNet's recommendations are indeed personalized for the target user and they could be \emph{explained} to a certain degree by tracing back to some entities in that specific user's explicit persona.



\section{Conclusion}

In this paper, we propose PerCoNet --- a novel deep persona-aware network with cross-view contrastive learning --- for personalized news recommendation. 
Different from the current mainstream approaches, PerCoNet builds an explicit persona for each user, and thus provides the news and user encoders with much richer and more personalized information. 
Additionally, the model includes an auxiliary task of cross-view contrastive learning, which further enhances its encoding capability.  
To the best of our knowledge, this is the first attempt to exploit explicit persona and also the first attempt to introduce contrastive learning in the context of personalized news recommendation.  
Extensive experiments on two real-world datasets, MIND and Adressa, have demonstrated that PerCoNet can significantly outperform state-of-the-art deep learning based competitors in recommending personalized news.
Furthermore, our ablation studies have shown that explicit persona and contrastive learning both contribute to PerCoNet's high recommendation performance. 



\bibliographystyle{ACM-Reference-Format}
\bibliography{refs}


\begin{thebibliography}{56}


\ifx \showCODEN    \undefined \def \showCODEN     #1{\unskip}     \fi
\ifx \showDOI      \undefined \def \showDOI       #1{#1}\fi
\ifx \showISBNx    \undefined \def \showISBNx     #1{\unskip}     \fi
\ifx \showISBNxiii \undefined \def \showISBNxiii  #1{\unskip}     \fi
\ifx \showISSN     \undefined \def \showISSN      #1{\unskip}     \fi
\ifx \showLCCN     \undefined \def \showLCCN      #1{\unskip}     \fi
\ifx \shownote     \undefined \def \shownote      #1{#1}          \fi
\ifx \showarticletitle \undefined \def \showarticletitle #1{#1}   \fi
\ifx \showURL      \undefined \def \showURL       {\relax}        \fi
\providecommand\bibfield[2]{#2}
\providecommand\bibinfo[2]{#2}
\providecommand\natexlab[1]{#1}
\providecommand\showeprint[2][]{arXiv:#2}

\bibitem[An et~al\mbox{.}(2019)]%
        {Mingxiao-LSTUR-ACL-2019}
\bibfield{author}{\bibinfo{person}{Mingxiao An}, \bibinfo{person}{Fangzhao Wu},
  \bibinfo{person}{Chuhan Wu}, \bibinfo{person}{Kun Zhang},
  \bibinfo{person}{Zheng Liu}, {and} \bibinfo{person}{Xing Xie}.}
  \bibinfo{year}{2019}\natexlab{}.
\newblock \showarticletitle{Neural News Recommendation with Long- and
  Short-term User Representations}. In \bibinfo{booktitle}{\emph{ACL}}.
  \bibinfo{pages}{336--345}.
\newblock


\bibitem[Bi et~al\mbox{.}(2022)]%
        {Qiwei-MTRec-ACL-2022}
\bibfield{author}{\bibinfo{person}{Qiwei Bi}, \bibinfo{person}{Jian Li},
  \bibinfo{person}{Lifeng Shang}, \bibinfo{person}{Xin Jiang},
  \bibinfo{person}{Qun Liu}, {and} \bibinfo{person}{Hanfang Yang}.}
  \bibinfo{year}{2022}\natexlab{}.
\newblock \showarticletitle{MTRec: Multi-Task Learning over {BERT} for News
  Recommendation}. In \bibinfo{booktitle}{\emph{ACL}}.
  \bibinfo{pages}{2663--2669}.
\newblock


\bibitem[Chen et~al\mbox{.}(2020)]%
        {Ting-SimCLE-ICML-2020}
\bibfield{author}{\bibinfo{person}{Ting Chen}, \bibinfo{person}{Simon
  Kornblith}, \bibinfo{person}{Mohammad Norouzi}, {and}
  \bibinfo{person}{Geoffrey~E. Hinton}.} \bibinfo{year}{2020}\natexlab{}.
\newblock \showarticletitle{A Simple Framework for Contrastive Learning of
  Visual Representations}. In \bibinfo{booktitle}{\emph{ICML}}.
  \bibinfo{pages}{1597--1607}.
\newblock


\bibitem[Choi et~al\mbox{.}(2022)]%
        {Seonghwan-Diversity-WWW-2022}
\bibfield{author}{\bibinfo{person}{Seonghwan Choi}, \bibinfo{person}{Hyeondey
  Kim}, {and} \bibinfo{person}{Manjun Gim}.} \bibinfo{year}{2022}\natexlab{}.
\newblock \showarticletitle{Do Not Read the Same News! Enhancing Diversity and
  Personalization of News Recommendation}. In \bibinfo{booktitle}{\emph{WWW}}.
  \bibinfo{pages}{1211--1215}.
\newblock


\bibitem[Devlin et~al\mbox{.}(2019)]%
        {Jacob-BERT-NAACL-2019}
\bibfield{author}{\bibinfo{person}{Jacob Devlin}, \bibinfo{person}{Ming{-}Wei
  Chang}, \bibinfo{person}{Kenton Lee}, {and} \bibinfo{person}{Kristina
  Toutanova}.} \bibinfo{year}{2019}\natexlab{}.
\newblock \showarticletitle{{BERT:} Pre-training of Deep Bidirectional
  Transformers for Language Understanding}. In
  \bibinfo{booktitle}{\emph{NAACL}}. \bibinfo{pages}{4171--4186}.
\newblock


\bibitem[Egozi et~al\mbox{.}(2011)]%
        {egozi2011concept}
\bibfield{author}{\bibinfo{person}{Ofer Egozi}, \bibinfo{person}{Shaul
  Markovitch}, {and} \bibinfo{person}{Evgeniy Gabrilovich}.}
  \bibinfo{year}{2011}\natexlab{}.
\newblock \showarticletitle{Concept-based Information Retrieval using Explicit
  Semantic Analysis}.
\newblock \bibinfo{journal}{\emph{ACM Transactions on Information Systems
  (TOIS)}} \bibinfo{volume}{29}, \bibinfo{number}{2} (\bibinfo{year}{2011}),
  \bibinfo{pages}{1--34}.
\newblock


\bibitem[Gabrilovich and Markovitch(2006)]%
        {gabrilovich2006overcoming}
\bibfield{author}{\bibinfo{person}{Evgeniy Gabrilovich} {and}
  \bibinfo{person}{Shaul Markovitch}.} \bibinfo{year}{2006}\natexlab{}.
\newblock \showarticletitle{Overcoming the Brittleness Bottleneck using
  {Wikipedia}: Enhancing Text Categorization with Encyclopedic Knowledge}. In
  \bibinfo{booktitle}{\emph{AAAI}}, Vol.~\bibinfo{volume}{6}.
  \bibinfo{pages}{1301--1306}.
\newblock


\bibitem[Gao et~al\mbox{.}(2021)]%
        {Tianyu-SimCSE-EMNLP-2021}
\bibfield{author}{\bibinfo{person}{Tianyu Gao}, \bibinfo{person}{Xingcheng
  Yao}, {and} \bibinfo{person}{Danqi Chen}.} \bibinfo{year}{2021}\natexlab{}.
\newblock \showarticletitle{SimCSE: Simple Contrastive Learning of Sentence
  Embeddings}. In \bibinfo{booktitle}{\emph{EMNLP}}.
  \bibinfo{pages}{6894--6910}.
\newblock


\bibitem[Giorgi et~al\mbox{.}(2021)]%
        {John-DeCLUTR-ACL/IJCNLP-2021}
\bibfield{author}{\bibinfo{person}{John~M. Giorgi}, \bibinfo{person}{Osvald
  Nitski}, \bibinfo{person}{Bo Wang}, {and} \bibinfo{person}{Gary~D. Bader}.}
  \bibinfo{year}{2021}\natexlab{}.
\newblock \showarticletitle{DeCLUTR: Deep Contrastive Learning for Unsupervised
  Textual Representations}. In \bibinfo{booktitle}{\emph{ACL/IJCNLP}}.
  \bibinfo{pages}{879--895}.
\newblock


\bibitem[Gong and Zhu(2022)]%
        {Shansan-Session-SIGIR-2022}
\bibfield{author}{\bibinfo{person}{Shansan Gong} {and}
  \bibinfo{person}{Kenny~Q. Zhu}.} \bibinfo{year}{2022}\natexlab{}.
\newblock \showarticletitle{Positive, Negative and Neutral: Modeling Implicit
  Feedback in Session-based News Recommendation}. In
  \bibinfo{booktitle}{\emph{SIGIR}}. \bibinfo{pages}{1185--1195}.
\newblock


\bibitem[Gulla et~al\mbox{.}(2017)]%
        {Jon-Adressa-WI-2017}
\bibfield{author}{\bibinfo{person}{Jon~Atle Gulla}, \bibinfo{person}{Lemei
  Zhang}, \bibinfo{person}{Peng Liu}, \bibinfo{person}{{\"{O}}zlem
  {\"{O}}zg{\"{o}}bek}, {and} \bibinfo{person}{Xiaomeng Su}.}
  \bibinfo{year}{2017}\natexlab{}.
\newblock \showarticletitle{The Adressa dataset for news recommendation}. In
  \bibinfo{booktitle}{\emph{WI}}. \bibinfo{pages}{1042--1048}.
\newblock


\bibitem[Guo et~al\mbox{.}(2017)]%
        {Huifeng-DeepFM-IJCAI-2017}
\bibfield{author}{\bibinfo{person}{Huifeng Guo}, \bibinfo{person}{Ruiming
  Tang}, \bibinfo{person}{Yunming Ye}, \bibinfo{person}{Zhenguo Li}, {and}
  \bibinfo{person}{Xiuqiang He}.} \bibinfo{year}{2017}\natexlab{}.
\newblock \showarticletitle{DeepFM: {A} Factorization-Machine based Neural
  Network for {CTR} Prediction}. In \bibinfo{booktitle}{\emph{IJCAI}}.
  \bibinfo{pages}{1725--1731}.
\newblock


\bibitem[He et~al\mbox{.}(2020)]%
        {Kaiming-Moco-CVPR-2020}
\bibfield{author}{\bibinfo{person}{Kaiming He}, \bibinfo{person}{Haoqi Fan},
  \bibinfo{person}{Yuxin Wu}, \bibinfo{person}{Saining Xie}, {and}
  \bibinfo{person}{Ross~B. Girshick}.} \bibinfo{year}{2020}\natexlab{}.
\newblock \showarticletitle{Momentum Contrast for Unsupervised Visual
  Representation Learning}. In \bibinfo{booktitle}{\emph{CVPR}}.
  \bibinfo{pages}{9726--9735}.
\newblock


\bibitem[Hu et~al\mbox{.}(2020)]%
        {Linmei-GNN-ACL-2020}
\bibfield{author}{\bibinfo{person}{Linmei Hu}, \bibinfo{person}{Siyong Xu},
  \bibinfo{person}{Chen Li}, \bibinfo{person}{Cheng Yang},
  \bibinfo{person}{Chuan Shi}, \bibinfo{person}{Nan Duan},
  \bibinfo{person}{Xing Xie}, {and} \bibinfo{person}{Ming Zhou}.}
  \bibinfo{year}{2020}\natexlab{}.
\newblock \showarticletitle{Graph Neural News Recommendation with Unsupervised
  Preference Disentanglement}. In \bibinfo{booktitle}{\emph{ACL}}.
  \bibinfo{pages}{4255--4264}.
\newblock


\bibitem[Joseph and Jiang(2019)]%
        {Kevin-KG-WWW-2019}
\bibfield{author}{\bibinfo{person}{Kevin Joseph} {and} \bibinfo{person}{Hui
  Jiang}.} \bibinfo{year}{2019}\natexlab{}.
\newblock \showarticletitle{Content based News Recommendation via Shortest
  Entity Distance over Knowledge Graphs}. In \bibinfo{booktitle}{\emph{WWW}}.
  \bibinfo{pages}{690--699}.
\newblock


\bibitem[Kazai et~al\mbox{.}(2016)]%
        {Gabriella-profile-SIGIR-2016}
\bibfield{author}{\bibinfo{person}{Gabriella Kazai}, \bibinfo{person}{Iskander
  Yusof}, {and} \bibinfo{person}{Daoud Clarke}.}
  \bibinfo{year}{2016}\natexlab{}.
\newblock \showarticletitle{Personalised News and Blog Recommendations based on
  User Location, Facebook and Twitter User Profiling}. In
  \bibinfo{booktitle}{\emph{SIGIR}}. \bibinfo{pages}{1129--1132}.
\newblock


\bibitem[Kingma and Ba(2015)]%
        {Diederik-Adam-ICLR-2015}
\bibfield{author}{\bibinfo{person}{Diederik~P. Kingma} {and}
  \bibinfo{person}{Jimmy Ba}.} \bibinfo{year}{2015}\natexlab{}.
\newblock \showarticletitle{Adam: {A} Method for Stochastic Optimization}. In
  \bibinfo{booktitle}{\emph{ICLR}}. \bibinfo{pages}{1--15}.
\newblock


\bibitem[Kipf et~al\mbox{.}(2020)]%
        {Thomas-WorldModels-ICLR-2020}
\bibfield{author}{\bibinfo{person}{Thomas~N. Kipf}, \bibinfo{person}{Elise
  van~der Pol}, {and} \bibinfo{person}{Max Welling}.}
  \bibinfo{year}{2020}\natexlab{}.
\newblock \showarticletitle{Contrastive Learning of Structured World Models}.
  In \bibinfo{booktitle}{\emph{ICLR}}. \bibinfo{pages}{1--18}.
\newblock


\bibitem[Ko et~al\mbox{.}(2022)]%
        {Ching-Revisiting-ICML-2022}
\bibfield{author}{\bibinfo{person}{Ching{-}Yun Ko}, \bibinfo{person}{Jeet
  Mohapatra}, \bibinfo{person}{Sijia Liu}, \bibinfo{person}{Pin{-}Yu Chen},
  \bibinfo{person}{Luca Daniel}, {and} \bibinfo{person}{Lily Weng}.}
  \bibinfo{year}{2022}\natexlab{}.
\newblock \showarticletitle{Revisiting Contrastive Learning through the Lens of
  Neighborhood Component Analysis: an Integrated Framework}. In
  \bibinfo{booktitle}{\emph{ICML}} \emph{(\bibinfo{series}{Proceedings of
  Machine Learning Research}, Vol.~\bibinfo{volume}{162})}.
  \bibinfo{pages}{11387--11412}.
\newblock


\bibitem[Li et~al\mbox{.}(2022)]%
        {Jian-Miner-ACL-2022}
\bibfield{author}{\bibinfo{person}{Jian Li}, \bibinfo{person}{Jieming Zhu},
  \bibinfo{person}{Qiwei Bi}, \bibinfo{person}{Guohao Cai},
  \bibinfo{person}{Lifeng Shang}, \bibinfo{person}{Zhenhua Dong},
  \bibinfo{person}{Xin Jiang}, {and} \bibinfo{person}{Qun Liu}.}
  \bibinfo{year}{2022}\natexlab{}.
\newblock \showarticletitle{{MINER:} Multi-Interest Matching Network for News
  Recommendation}. In \bibinfo{booktitle}{\emph{ACL}}.
  \bibinfo{pages}{343--352}.
\newblock


\bibitem[Lian et~al\mbox{.}(2018)]%
        {Jianxun-MultiChannel-IJCAI-2018}
\bibfield{author}{\bibinfo{person}{Jianxun Lian}, \bibinfo{person}{Fuzheng
  Zhang}, \bibinfo{person}{Xing Xie}, {and} \bibinfo{person}{Guangzhong Sun}.}
  \bibinfo{year}{2018}\natexlab{}.
\newblock \showarticletitle{Towards Better Representation Learning for News
  Recommendation: a Multi-Channel Deep Fusion Approach}. In
  \bibinfo{booktitle}{\emph{IJCAI}}. \bibinfo{pages}{3805--3811}.
\newblock


\bibitem[Liu et~al\mbox{.}(2021)]%
        {Danyang-anchorGK-KDD-2021}
\bibfield{author}{\bibinfo{person}{Danyang Liu}, \bibinfo{person}{Jianxun
  Lian}, \bibinfo{person}{Zheng Liu}, \bibinfo{person}{Xiting Wang},
  \bibinfo{person}{Guangzhong Sun}, {and} \bibinfo{person}{Xing Xie}.}
  \bibinfo{year}{2021}\natexlab{}.
\newblock \showarticletitle{Reinforced Anchor Knowledge Graph Generation for
  News Recommendation Reasoning}. In \bibinfo{booktitle}{\emph{KDD}}.
  \bibinfo{pages}{1055--1065}.
\newblock


\bibitem[Liu et~al\mbox{.}(2020a)]%
        {Danyang-KRED-RecSys-2020}
\bibfield{author}{\bibinfo{person}{Danyang Liu}, \bibinfo{person}{Jianxun
  Lian}, \bibinfo{person}{Shiyin Wang}, \bibinfo{person}{Ying Qiao},
  \bibinfo{person}{Jiun{-}Hung Chen}, \bibinfo{person}{Guangzhong Sun}, {and}
  \bibinfo{person}{Xing Xie}.} \bibinfo{year}{2020}\natexlab{a}.
\newblock \showarticletitle{{KRED:} Knowledge-Aware Document Representation for
  News Recommendations}. In \bibinfo{booktitle}{\emph{RecSys}}.
  \bibinfo{pages}{200--209}.
\newblock


\bibitem[Liu et~al\mbox{.}(2020b)]%
        {Yong-HyperNews-IJCAI-2020}
\bibfield{author}{\bibinfo{person}{Rui Liu}, \bibinfo{person}{Huilin Peng},
  \bibinfo{person}{Yong Chen}, {and} \bibinfo{person}{Dell Zhang}.}
  \bibinfo{year}{2020}\natexlab{b}.
\newblock \showarticletitle{HyperNews: Simultaneous News Recommendation and
  Active-Time Prediction via a Double-Task Deep Neural Network}. In
  \bibinfo{booktitle}{\emph{IJCAI}}. \bibinfo{pages}{3487--3493}.
\newblock


\bibitem[Liu et~al\mbox{.}(2019)]%
        {Yinhan-RoBERTa-CoRR-2019}
\bibfield{author}{\bibinfo{person}{Yinhan Liu}, \bibinfo{person}{Myle Ott},
  \bibinfo{person}{Naman Goyal}, \bibinfo{person}{Jingfei Du},
  \bibinfo{person}{Mandar Joshi}, \bibinfo{person}{Danqi Chen},
  \bibinfo{person}{Omer Levy}, \bibinfo{person}{Mike Lewis},
  \bibinfo{person}{Luke Zettlemoyer}, {and} \bibinfo{person}{Veselin
  Stoyanov}.} \bibinfo{year}{2019}\natexlab{}.
\newblock \showarticletitle{RoBERTa: {A} Robustly Optimized {BERT} Pretraining
  Approach}.
\newblock \bibinfo{journal}{\emph{CoRR}}  \bibinfo{volume}{abs/1907.11692}
  (\bibinfo{year}{2019}), \bibinfo{pages}{1--13}.
\newblock


\bibitem[Ma et~al\mbox{.}(2016)]%
        {Hao-Fatigue-WWW-2016}
\bibfield{author}{\bibinfo{person}{Hao Ma}, \bibinfo{person}{Xueqing Liu},
  {and} \bibinfo{person}{Zhihong Shen}.} \bibinfo{year}{2016}\natexlab{}.
\newblock \showarticletitle{User Fatigue in Online News Recommendation}. In
  \bibinfo{booktitle}{\emph{WWW}}. \bibinfo{pages}{1363--1372}.
\newblock


\bibitem[Meng et~al\mbox{.}(2021)]%
        {Lingkang-DCAN-DASFAA-2021}
\bibfield{author}{\bibinfo{person}{Lingkang Meng}, \bibinfo{person}{Chongyang
  Shi}, \bibinfo{person}{Shufeng Hao}, {and} \bibinfo{person}{Xiangrui Su}.}
  \bibinfo{year}{2021}\natexlab{}.
\newblock \showarticletitle{{DCAN:} Deep Co-Attention Network by Modeling User
  Preference and News Lifecycle for News Recommendation}. In
  \bibinfo{booktitle}{\emph{DASFAA}}. \bibinfo{pages}{100--114}.
\newblock


\bibitem[Peng et~al\mbox{.}(2021)]%
        {Jizong-SelfPaced-NeurIPS-2021}
\bibfield{author}{\bibinfo{person}{Jizong Peng}, \bibinfo{person}{Ping Wang},
  \bibinfo{person}{Christian Desrosiers}, {and} \bibinfo{person}{Marco
  Pedersoli}.} \bibinfo{year}{2021}\natexlab{}.
\newblock \showarticletitle{Self-Paced Contrastive Learning for Semi-supervised
  Medical Image Segmentation with Meta-labels}. In
  \bibinfo{booktitle}{\emph{NeurIPS}}. \bibinfo{pages}{16686--16699}.
\newblock


\bibitem[Purushwalkam and Gupta(2020)]%
        {Senthil-DemystifyingCL-NeurIPS-2020}
\bibfield{author}{\bibinfo{person}{Senthil Purushwalkam} {and}
  \bibinfo{person}{Abhinav Gupta}.} \bibinfo{year}{2020}\natexlab{}.
\newblock \showarticletitle{Demystifying Contrastive Self-Supervised Learning:
  Invariances, Augmentations and Dataset Biases}. In
  \bibinfo{booktitle}{\emph{NeurIPS}}.
\newblock


\bibitem[Qiu et~al\mbox{.}(2022)]%
        {Zhaopeng-GREP-TKDD-2022}
\bibfield{author}{\bibinfo{person}{Zhaopeng Qiu}, \bibinfo{person}{Yunfan Hu},
  {and} \bibinfo{person}{Xian Wu}.} \bibinfo{year}{2022}\natexlab{}.
\newblock \showarticletitle{Graph Neural News Recommendation with User Existing
  and Potential Interest Modeling}.
\newblock \bibinfo{journal}{\emph{ACM Transactions on Knowledge Discovery from
  Data}} \bibinfo{volume}{16}, \bibinfo{number}{5} (\bibinfo{year}{2022}),
  \bibinfo{pages}{96:1--96:17}.
\newblock


\bibitem[Rendle(2012)]%
        {Steffen-LibFM-TIST-2012}
\bibfield{author}{\bibinfo{person}{Steffen Rendle}.}
  \bibinfo{year}{2012}\natexlab{}.
\newblock \showarticletitle{Factorization Machines with libFM}.
\newblock \bibinfo{journal}{\emph{ACM Transactions on Intelligent Systems and
  Technology}} \bibinfo{volume}{3}, \bibinfo{number}{3} (\bibinfo{year}{2012}),
  \bibinfo{pages}{57:1--57:22}.
\newblock


\bibitem[Shi et~al\mbox{.}(2021)]%
        {Shaoyun-WG4Rec-CIKM-2021}
\bibfield{author}{\bibinfo{person}{Shaoyun Shi}, \bibinfo{person}{Weizhi Ma},
  \bibinfo{person}{Zhen Wang}, \bibinfo{person}{Min Zhang},
  \bibinfo{person}{Kun Fang}, \bibinfo{person}{Jingfang Xu},
  \bibinfo{person}{Yiqun Liu}, {and} \bibinfo{person}{Shaoping Ma}.}
  \bibinfo{year}{2021}\natexlab{}.
\newblock \showarticletitle{WG4Rec: Modeling Textual Content with Word Graph
  for News Recommendation}. In \bibinfo{booktitle}{\emph{CIKM}}.
  \bibinfo{pages}{1651--1660}.
\newblock


\bibitem[Sun et~al\mbox{.}(2022)]%
        {Tiening-Rumor-WWW-2022}
\bibfield{author}{\bibinfo{person}{Tiening Sun}, \bibinfo{person}{Zhong Qian},
  \bibinfo{person}{Sujun Dong}, \bibinfo{person}{Peifeng Li}, {and}
  \bibinfo{person}{Qiaoming Zhu}.} \bibinfo{year}{2022}\natexlab{}.
\newblock \showarticletitle{Rumor Detection on Social Media with Graph
  Adversarial Contrastive Learning}. In \bibinfo{booktitle}{\emph{WWW}}.
  \bibinfo{publisher}{{ACM}}, \bibinfo{pages}{2789--2797}.
\newblock


\bibitem[Tian et~al\mbox{.}(2020)]%
        {Yonglong-goodCL-NeurIPS-2020}
\bibfield{author}{\bibinfo{person}{Yonglong Tian}, \bibinfo{person}{Chen Sun},
  \bibinfo{person}{Ben Poole}, \bibinfo{person}{Dilip Krishnan},
  \bibinfo{person}{Cordelia Schmid}, {and} \bibinfo{person}{Phillip Isola}.}
  \bibinfo{year}{2020}\natexlab{}.
\newblock \showarticletitle{What Makes for Good Views for Contrastive
  Learning?}. In \bibinfo{booktitle}{\emph{NeurIPS}}.
\newblock


\bibitem[Tian et~al\mbox{.}(2021)]%
        {Yu-ReGCN-SIGIR-2021}
\bibfield{author}{\bibinfo{person}{Yu Tian}, \bibinfo{person}{Yuhao Yang},
  \bibinfo{person}{Xudong Ren}, \bibinfo{person}{Pengfei Wang},
  \bibinfo{person}{Fangzhao Wu}, \bibinfo{person}{Qian Wang}, {and}
  \bibinfo{person}{Chenliang Li}.} \bibinfo{year}{2021}\natexlab{}.
\newblock \showarticletitle{Joint Knowledge Pruning and Recurrent Graph
  Convolution for News Recommendation}. In \bibinfo{booktitle}{\emph{SIGIR}}.
  \bibinfo{pages}{51--60}.
\newblock


\bibitem[Tong et~al\mbox{.}(2021)]%
        {Xiaohai-RAP-IJCAI-2021}
\bibfield{author}{\bibinfo{person}{Xiaohai Tong}, \bibinfo{person}{Pengfei
  Wang}, \bibinfo{person}{Chenliang Li}, \bibinfo{person}{Long Xia}, {and}
  \bibinfo{person}{ShaoZhang Niu}.} \bibinfo{year}{2021}\natexlab{}.
\newblock \showarticletitle{Pattern-enhanced Contrastive Policy Learning
  Network for Sequential Recommendation}. In \bibinfo{booktitle}{\emph{IJCAI}}.
  \bibinfo{pages}{1593--1599}.
\newblock


\bibitem[Tran et~al\mbox{.}(2021)]%
        {Dai-CUPMAR-WISE-2021}
\bibfield{author}{\bibinfo{person}{Dai~Hoang Tran},
  \bibinfo{person}{Salma~Abdalla Hamad}, \bibinfo{person}{Munazza Zaib}, {and}
  \bibinfo{person}{et. al.}} \bibinfo{year}{2021}\natexlab{}.
\newblock \showarticletitle{Deep News Recommendation with Contextual User
  Profiling and Multifaceted Article Representation}. In
  \bibinfo{booktitle}{\emph{WISE}}. \bibinfo{pages}{237--251}.
\newblock


\bibitem[Vaswani et~al\mbox{.}(2017)]%
        {Ashish-AttentionAllNeed-NeurIPS-2017}
\bibfield{author}{\bibinfo{person}{Ashish Vaswani}, \bibinfo{person}{Noam
  Shazeer}, \bibinfo{person}{Niki Parmar}, \bibinfo{person}{Jakob Uszkoreit},
  \bibinfo{person}{Llion Jones}, \bibinfo{person}{Aidan~N. Gomez},
  \bibinfo{person}{Lukasz Kaiser}, {and} \bibinfo{person}{Illia Polosukhin}.}
  \bibinfo{year}{2017}\natexlab{}.
\newblock \showarticletitle{Attention is All you Need}. In
  \bibinfo{booktitle}{\emph{NeurIPS}}. \bibinfo{pages}{5998--6008}.
\newblock


\bibitem[Wang et~al\mbox{.}(2020)]%
        {Heyuan-FIM-ACL-2020}
\bibfield{author}{\bibinfo{person}{Heyuan Wang}, \bibinfo{person}{Fangzhao Wu},
  \bibinfo{person}{Zheng Liu}, {and} \bibinfo{person}{Xing Xie}.}
  \bibinfo{year}{2020}\natexlab{}.
\newblock \showarticletitle{Fine-grained Interest Matching for Neural News
  Recommendation}. In \bibinfo{booktitle}{\emph{ACL}}.
  \bibinfo{pages}{836--845}.
\newblock


\bibitem[Wang et~al\mbox{.}(2018)]%
        {Hongwei-DKN-WWW-2018}
\bibfield{author}{\bibinfo{person}{Hongwei Wang}, \bibinfo{person}{Fuzheng
  Zhang}, \bibinfo{person}{Xing Xie}, {and} \bibinfo{person}{Minyi Guo}.}
  \bibinfo{year}{2018}\natexlab{}.
\newblock \showarticletitle{{DKN:} Deep Knowledge-Aware Network for News
  Recommendation}. In \bibinfo{booktitle}{\emph{WWW}}.
  \bibinfo{pages}{1835--1844}.
\newblock


\bibitem[Wang et~al\mbox{.}(2021)]%
        {Jingkun-PENR-CIKM-2021}
\bibfield{author}{\bibinfo{person}{Jingkun Wang}, \bibinfo{person}{Yipu Chen},
  \bibinfo{person}{Zichun Wang}, {and} \bibinfo{person}{Wen Zhao}.}
  \bibinfo{year}{2021}\natexlab{}.
\newblock \showarticletitle{Popularity-Enhanced News Recommendation with
  Multi-View Interest Representation}. In \bibinfo{booktitle}{\emph{CIKM}}.
  \bibinfo{pages}{1949--1958}.
\newblock


\bibitem[Wang et~al\mbox{.}(2022)]%
        {Rongyao-MINS-ICASSP-2022}
\bibfield{author}{\bibinfo{person}{Rongyao Wang}, \bibinfo{person}{Shoujin
  Wang}, \bibinfo{person}{Wenpeng Lu}, {and} \bibinfo{person}{Xueping Peng}.}
  \bibinfo{year}{2022}\natexlab{}.
\newblock \showarticletitle{News Recommendation Via Multi-Interest News
  Sequence Modelling}. In \bibinfo{booktitle}{\emph{ICASSP}}.
  \bibinfo{publisher}{{IEEE}}, \bibinfo{pages}{7942--7946}.
\newblock


\bibitem[Wu et~al\mbox{.}(2019a)]%
        {Chuhan-NAML-IJCAI-2019}
\bibfield{author}{\bibinfo{person}{Chuhan Wu}, \bibinfo{person}{Fangzhao Wu},
  \bibinfo{person}{Mingxiao An}, \bibinfo{person}{Jianqiang Huang},
  \bibinfo{person}{Yongfeng Huang}, {and} \bibinfo{person}{Xing Xie}.}
  \bibinfo{year}{2019}\natexlab{a}.
\newblock \showarticletitle{Neural News Recommendation with Attentive
  Multi-View Learning}. In \bibinfo{booktitle}{\emph{IJCAI}}.
  \bibinfo{pages}{3863--3869}.
\newblock


\bibitem[Wu et~al\mbox{.}(2019b)]%
        {Chuhan-NRMS-EMNLP-2019}
\bibfield{author}{\bibinfo{person}{Chuhan Wu}, \bibinfo{person}{Fangzhao Wu},
  \bibinfo{person}{Mingxiao An}, \bibinfo{person}{Jianqiang Huang},
  \bibinfo{person}{Yongfeng Huang}, {and} \bibinfo{person}{Xing Xie}.}
  \bibinfo{year}{2019}\natexlab{b}.
\newblock \showarticletitle{Neural News Recommendation with Multi-Head
  Self-Attention}. In \bibinfo{booktitle}{\emph{EMNLP}}.
  \bibinfo{pages}{3863--3869}.
\newblock


\bibitem[Wu et~al\mbox{.}(2019c)]%
        {Chuhan-NPA-KDD-2019}
\bibfield{author}{\bibinfo{person}{Chuhan Wu}, \bibinfo{person}{Fangzhao Wu},
  \bibinfo{person}{Mingxiao An}, \bibinfo{person}{Jianqiang Huang},
  \bibinfo{person}{Yongfeng Huang}, {and} \bibinfo{person}{Xing Xie}.}
  \bibinfo{year}{2019}\natexlab{c}.
\newblock \showarticletitle{{NPA:} Neural News Recommendation with Personalized
  Attention}. In \bibinfo{booktitle}{\emph{KDD}}. \bibinfo{pages}{2576--2584}.
\newblock


\bibitem[Wu et~al\mbox{.}(2021)]%
        {Chuhan-User-as-Graph-IJCAI-2021}
\bibfield{author}{\bibinfo{person}{Chuhan Wu}, \bibinfo{person}{Fangzhao Wu},
  \bibinfo{person}{Yongfeng Huang}, {and} \bibinfo{person}{Xing Xie}.}
  \bibinfo{year}{2021}\natexlab{}.
\newblock \showarticletitle{User-as-Graph: User Modeling with Heterogeneous
  Graph Pooling for News Recommendation}. In \bibinfo{booktitle}{\emph{IJCAI}}.
  \bibinfo{pages}{1624--1630}.
\newblock


\bibitem[Wu et~al\mbox{.}(2020b)]%
        {Chuhan-CPRS-IJCAI-2020}
\bibfield{author}{\bibinfo{person}{Chuhan Wu}, \bibinfo{person}{Fangzhao Wu},
  \bibinfo{person}{Tao Qi}, {and} \bibinfo{person}{Yongfeng Huang}.}
  \bibinfo{year}{2020}\natexlab{b}.
\newblock \showarticletitle{User Modeling with Click Preference and Reading
  Satisfaction for News Recommendation}. In \bibinfo{booktitle}{\emph{IJCAI}}.
  \bibinfo{pages}{3023--3029}.
\newblock


\bibitem[Wu et~al\mbox{.}(2022a)]%
        {Chuhan-FeedRec-WWW-2022}
\bibfield{author}{\bibinfo{person}{Chuhan Wu}, \bibinfo{person}{Fangzhao Wu},
  \bibinfo{person}{Tao Qi}, \bibinfo{person}{Qi Liu}, \bibinfo{person}{Xuan
  Tian}, \bibinfo{person}{Jie Li}, \bibinfo{person}{Wei He},
  \bibinfo{person}{Yongfeng Huang}, {and} \bibinfo{person}{Xing Xie}.}
  \bibinfo{year}{2022}\natexlab{a}.
\newblock \showarticletitle{FeedRec: News Feed Recommendation with Various User
  Feedbacks}. In \bibinfo{booktitle}{\emph{WWW}}. \bibinfo{pages}{2088--2097}.
\newblock


\bibitem[Wu et~al\mbox{.}(2022b)]%
        {Chuhan-MMRec-SIGIR-2022}
\bibfield{author}{\bibinfo{person}{Chuhan Wu}, \bibinfo{person}{Fangzhao Wu},
  \bibinfo{person}{Tao Qi}, \bibinfo{person}{Chao Zhang},
  \bibinfo{person}{Yongfeng Huang}, {and} \bibinfo{person}{Tong Xu}.}
  \bibinfo{year}{2022}\natexlab{b}.
\newblock \showarticletitle{MM-Rec: Visiolinguistic Model Empowered Multimodal
  News Recommendation}. In \bibinfo{booktitle}{\emph{SIGIR}}.
  \bibinfo{pages}{2560--2564}.
\newblock


\bibitem[Wu et~al\mbox{.}(2020a)]%
        {Fangzhao-MIND-ACL-2020}
\bibfield{author}{\bibinfo{person}{Fangzhao Wu}, \bibinfo{person}{Ying Qiao},
  \bibinfo{person}{Jiun{-}Hung Chen}, \bibinfo{person}{Chuhan Wu},
  \bibinfo{person}{Tao Qi}, \bibinfo{person}{Jianxun Lian},
  \bibinfo{person}{Danyang Liu}, \bibinfo{person}{Xing Xie},
  \bibinfo{person}{Jianfeng Gao}, \bibinfo{person}{Winnie Wu}, {and}
  \bibinfo{person}{Ming Zhou}.} \bibinfo{year}{2020}\natexlab{a}.
\newblock \showarticletitle{{MIND:} {A} Large-scale Dataset for News
  Recommendation}. In \bibinfo{booktitle}{\emph{ACL}}.
  \bibinfo{pages}{3597--3606}.
\newblock


\bibitem[Xia et~al\mbox{.}(2021)]%
        {Xin-DHCN-AAAI-2021}
\bibfield{author}{\bibinfo{person}{Xin Xia}, \bibinfo{person}{Hongzhi Yin},
  \bibinfo{person}{Junliang Yu}, \bibinfo{person}{Qinyong Wang},
  \bibinfo{person}{Lizhen Cui}, {and} \bibinfo{person}{Xiangliang Zhang}.}
  \bibinfo{year}{2021}\natexlab{}.
\newblock \showarticletitle{Self-Supervised Hypergraph Convolutional Networks
  for Session-based Recommendation}. In \bibinfo{booktitle}{\emph{AAAI}}.
  \bibinfo{pages}{4503--4511}.
\newblock


\bibitem[Yi et~al\mbox{.}(2021)]%
        {Jingwei-Federated-EMNLP-2021}
\bibfield{author}{\bibinfo{person}{Jingwei Yi}, \bibinfo{person}{Fangzhao Wu},
  \bibinfo{person}{Chuhan Wu}, {and} \bibinfo{person}{et. al.}}
  \bibinfo{year}{2021}\natexlab{}.
\newblock \showarticletitle{Efficient-FedRec: Efficient Federated Learning
  Framework for Privacy-Preserving News Recommendation}. In
  \bibinfo{booktitle}{\emph{EMNLP}}. \bibinfo{pages}{2814--2824}.
\newblock


\bibitem[Zhang et~al\mbox{.}(2021a)]%
        {Qi-UNBERT-IJCAI-2021}
\bibfield{author}{\bibinfo{person}{Qi Zhang}, \bibinfo{person}{Jingjie Li},
  \bibinfo{person}{Qinglin Jia}, \bibinfo{person}{Chuyuan Wang},
  \bibinfo{person}{Jieming Zhu}, \bibinfo{person}{Zhaowei Wang}, {and}
  \bibinfo{person}{Xiuqiang He}.} \bibinfo{year}{2021}\natexlab{a}.
\newblock \showarticletitle{{UNBERT:} User-News Matching {BERT} for News
  Recommendation}. In \bibinfo{booktitle}{\emph{IJCAI}}.
  \bibinfo{pages}{3356--3362}.
\newblock


\bibitem[Zhang et~al\mbox{.}(2021b)]%
        {Xuanyu-EEG-WWW-2021}
\bibfield{author}{\bibinfo{person}{Xuanyu Zhang}, \bibinfo{person}{Qing Yang},
  {and} \bibinfo{person}{Dongliang Xu}.} \bibinfo{year}{2021}\natexlab{b}.
\newblock \showarticletitle{Combining Explicit Entity Graph with Implicit Text
  Information for News Recommendation}. In \bibinfo{booktitle}{\emph{WWW}}.
  \bibinfo{pages}{412--416}.
\newblock


\bibitem[Zhu et~al\mbox{.}(2019)]%
        {Qiannan-DAN-AAAI-2019}
\bibfield{author}{\bibinfo{person}{Qiannan Zhu}, \bibinfo{person}{Xiaofei
  Zhou}, \bibinfo{person}{Zeliang Song}, \bibinfo{person}{Jianlong Tan}, {and}
  \bibinfo{person}{Li Guo}.} \bibinfo{year}{2019}\natexlab{}.
\newblock \showarticletitle{{DAN:} Deep Attention Neural Network for News
  Recommendation}. In \bibinfo{booktitle}{\emph{AAAI}}.
  \bibinfo{pages}{5973--5980}.
\newblock


\bibitem[Zolfaghari et~al\mbox{.}(2021)]%
        {Mohammadreza-CrossCLR-ICCV-2021}
\bibfield{author}{\bibinfo{person}{Mohammadreza Zolfaghari},
  \bibinfo{person}{Yi Zhu}, \bibinfo{person}{Peter~V. Gehler}, {and}
  \bibinfo{person}{Thomas Brox}.} \bibinfo{year}{2021}\natexlab{}.
\newblock \showarticletitle{CrossCLR: Cross-modal Contrastive Learning For
  Multi-modal Video Representations}. In \bibinfo{booktitle}{\emph{ICCV}}.
  \bibinfo{pages}{1430--1439}.
\newblock


\end{thebibliography}

%
%
%
%
%
%
%
%

\end{document}